\documentclass[12pt,a4paper]{article}
\usepackage[a4paper,left = 2.5cm, right = 2.5cm, top = 2.4cm, bottom = 4cm]{geometry}
\usepackage{mathptmx} 
\usepackage{avant}     
\usepackage[affil-it]{authblk}
\usepackage[english]{babel}
\usepackage{float}

\usepackage{latexsym,amsfonts,amssymb}
\usepackage{amssymb,amsmath,bm,mathrsfs,makeidx,amsfonts,graphicx,amsthm}
\usepackage[authoryear]{natbib}
\usepackage{epsfig}
\usepackage{setspace}
\usepackage{subfig}
\usepackage{graphicx}
\usepackage{graphics}
\usepackage{booktabs}
\usepackage{multirow}
\usepackage{lscape}
\usepackage{pdflscape}
\usepackage{rotating}
\usepackage{multirow}
\usepackage{indentfirst}
\usepackage{caption}
\usepackage{array}
\usepackage{xcolor}
\usepackage{mathtools}
\usepackage{ltxtable}

\usepackage[colorlinks=true,citecolor=blue]{hyperref}

\usepackage{tabularx,booktabs}
\newcolumntype{Y}{>{\centering\arraybackslash}X}

\newtheorem{rmk}{Remark}

\numberwithin{equation}{section}
\def\ba{\begin{align*}}
\def\ea{\end{align*}}
\def\bao{\begin{align}}
\def\eao{\end{align}}
\def\begine{\begin{enumerate}}
\def\ende{\end{enumerate}}

\def\be{\begin{equation}}
\def\ee{\end{equation}}

\DeclareMathOperator{\Tr}{Tr}

\newcommand{\bbX}{{\boldsymbol X}}

\newcommand{\bbB}{{\boldsymbol B}}

\newcommand{\bbb}{{\boldsymbol b}}

\newcommand{\bbH}{{\boldsymbol H}}

\newcommand{\bbk}{{\boldsymbol k}}
\newcommand{\bbK}{{\boldsymbol K}}

\newcommand{\bbM}{{\boldsymbol M}}

\newcommand{\bbW}{{\boldsymbol W}}

\newcommand{\bvar}{\boldsymbol{\varepsilon}}

\newcommand{\RN}[1]{%
  \textup{\uppercase\expandafter{\romannumeral#1}}%
}

\begin{document}
\title{Mortality Forecasting using Factor Models: Time-varying or Time-invariant Factor Loadings?
} 
\author[1]{Lingyu He}
\author[2]{Fei Huang\thanks{Correspondence to: Dr. Fei Huang,  School of Risk and Actuarial Studies, UNSW Business School, UNSW Sydney, NSW 2052, Australia. Email: feihuang@unsw.edu.au}}
\author[3]{Jianjie Shi}
\author[4]{Yanrong Yang}
\affil[1]{Hunan University}
\affil[2]{The University of New South Wales}
\affil[3]{Monash University}
\affil[4]{The Australian National University}
\date{}
\maketitle

\begin{abstract}
	
Many existing mortality models follow the framework of classical factor models, such as the Lee-Carter model and its variants. Latent common factors in factor models are defined as  time-related mortality indices (such as $\kappa_t$ in the Lee-Carter model). Factor loadings, which capture the linear relationship between age variables and latent common factors (such as $\beta_x$ in the Lee-Carter model), are assumed to be time-invariant in the classical framework. This assumption is usually too restrictive in reality as mortality datasets typically span a long period of time. Driving forces such as medical improvement of certain diseases, environmental changes and technological progress may significantly influence the relationship of different variables. In this paper, we first develop a factor model with time-varying factor loadings (time-varying factor model) as an extension of the classical factor model for mortality modelling. 
Two forecasting methods to extrapolate the factor loadings, the local regression method and the naive method, are proposed for the time-varying factor model. From the empirical data analysis, we find that the new model can capture the empirical feature of time-varying factor loadings and  improve mortality forecasting over different horizons and countries. Further, we propose a novel approach based on change point analysis to estimate the optimal `boundary' between short-term and long-term forecasting, which is favoured by the local linear regression and naive method, respectively. Additionally, simulation studies are provided to show the performance of the time-varying factor model under various scenarios.    


\end{abstract}
\providecommand{\jelcode}[1]{\textbf{JEL Code\ \ \ } #1}
\providecommand{\keywords}[1]{\textbf{Keywords\ \ \ } #1}

\jelcode{G22 - Insurance; Insurance Companies; Actuarial Studies}


\section{Introduction}
\label{sec: intro}
Mortality forecasting is an important topic in various areas, such as demography, actuarial science and government policymaking. Most age-specific mortality data are high-dimensional time series. 
The factor model approach is one of the most popular methods to model high-dimensional time series, representing the data matrix by a few latent common factors. Common factors describe common information shared by cross-sections, while factor loadings reflect the linear relationship between the original variables and the common factors. 
There is a large literature discussing factor models, including but not limited to \cite{Anderson1963}, \cite{pena1987identifying}, \cite{stock2002forecasting}, \cite{bai2002determining}, \cite{bai2009panel}, \cite{lam2012factor} and \cite{chang2018principal}. 

Many existing stochastic mortality models use the factor model approach.  As an application of the classical factor model (with time-invariant factor loadings), \cite{lee1992modeling} (Lee-Carter Model) is one of the most prominent methods for mortality forecasting, which is employed by the US Bureau of the Census as the benchmark model to predict long-run life expectancy \citep{hollmann1999methodology}. The common factor extracted by the Lee-Carter model is defined as Mortality Index, and the factor loadings capture the relationship between the age variables and the mortality index. Since there is only one factor in the Lee-Carter model,  \cite{booth2002applying}, \cite{renshaw2003lee} and \cite{yang2010modeling} extended the Lee-Carter framework  to incorporate more common latent factors for mortality modelling in different countries.  \cite{li2005outlier} proposed an outlier-adjusted model to deal with possible outliers in the mortality index by combining the Lee-Carter model with time series outlier analysis. Additionally, \cite{booth2006lee} compared the Lee-Carter model with four other variants by applying them to mortality data of multiple populations. \cite{tuljapurkar2000universal} examined mortality rates over five decades for the G7 countries using the Lee-Carter model. \cite{lundstrom2004mortality} and \cite{booth2004beyond} applied the Lee-Carter model to mortality data of Sweden and Australia, respectively. A summary of the variants of the Lee-Carter model is discussed in \cite{booth2008mortality}.

In the existing literature of mortality factor models, factor loadings, which capture the relationship between age variables and latent common factors, are usually assumed to be invariant over time (we call factor models with time-invariant factor loadings `classical factor model'). For example, in Lee-Carter model, there is only one factor and the time-invariant factor loading represents the age-related sensitivity to the mortality improvement (we call classical factor model with only one factor `Lee-Carter model' throughout this paper). However, since mortality datasets typically span a long period of time, it is restrictive to assume that the factor loadings are time-invariant. Driving forces such as medical improvement of certain diseases, environmental changes, and technological progress may influence the relationship of different variables significantly. \citet{booth2002applying} studied the violation of the invariance assumption in the mortality data of Australia and suggested to find an optimal fitting period during which the factor loadings were invariant to improve the fit of the Lee-Carter model. Their approach, however, needs to manually select the fitting period and hence loses the information of early years. In recent years, there is a rich literature on time-varying factor models to capture the dynamics and structural changes in factor loadings for macroeconomic variables modelling, for example, see \cite{BREITUNG201171} and \cite{CHEN201430}.  However, there has been no literature on mortality modelling which allows factor loadings to change smoothly over time, to the best of our knowledge. \cite{LI2017166} and \cite{LI2015264} used semi-parametric approaches to extend the CBD models \citep{Cairns2009} by allowing for time-varying coefficients, which can free model assumptions and show superior short-term forecasting performance. However, CBD models are only suitable for old-age mortality modelling, and the factors (regressors) are observable. 
Unfortunately, for Lee-Carter model and many of its variants the factors are unobserved, which makes it difficult to model and estimate. To fill those gaps,  we introduce a  factor model with time-varying factor loadings as an extension of the classical factor model based on \cite{su2017time}. This new model can be used for mortality modelling and forecasting by developing corresponding estimation and forecasting methods. 

As the time-varying factor model allows for time-varying factor loadings, it provides more flexibility in model fitting, which, however, also poses challenges in model forecasting. Besides forecasting the common factors, factor loadings also need to be extrapolated into the future. In this paper, we provide two forecasting methods of the factor loadings, one uses the local linear regression to roll over the time-varying factor loadings into the future; while the other one inherits the value of the factor loading from the last time period and remains invariant in the future. These two forecasting methods are called the local linear regression and the naive method, respectively. Their details are described in Section \ref{sec: model}. Empirical results using the mortality data from different populations show that the time-varying factor model provides more accurate out-of-sample forecasting results than the Lee-Carter model.

The existing literature suggests that different forecasting horizons may favour different models. For example, \citet{bell1997comparing} found that a simple random walk with drift model for age-specific mortality rates yields the most accurate 1-step-ahead forecast compared with the other six methods on the US data. \citet{hyndman2007robust} introduced a method which outperformed the method proposed by \citet{lee2001evaluating} in the long-term forecasting.  Specifically, we have found in the literature that semi-parametric or non-parametric methods can be more suitable for short-term forecasting. For example, the semi-parametric model developed in \cite{LI2015264} can produce superior 5-year-ahead forecasting results. \cite{CMI2009} employed the P-splines model (\cite{currie2004smoothing}) for short-term forecasting to generate the initial rates of mortality improvement. Our empirical applications in Section \ref{subsec: comparison} also suggest that the time-varying model based on local regression (non-parametric forecasting) is better for short-term forecasting, while the time-varying model based on naive method (parametric forecasting) is better for long-term forecasting. Then where is the optimal `boundary' between short-term (based on the local regression method) and long-term (based on the naive method) forecasting? We propose a novel approach based on change point analysis \citep{bai2010common} to estimate the optimal `boundary' and apply it to mortality data of multiple countries. Additionally, we conduct simulation studies to show the performance of the time-varying factor model under different scenarios and investigate under which conditions it preforms better than the classical factor model.


The rest of the paper is organized as follows. Section \ref{sec: model} introduces the time-varying factor model and its estimation approach. The forecasting methods based on the time-varying factor model are also discussed in detail. Section \ref{sec: choice} discusses the relative advantages  of the local regression method and the naive method in the short-term and long-term forecasting, respectively. We then propose an approach based on change point analysis to estimate the `boundary' between short-term and long-term forecasting, which is favoured by the local regression method and the naive method, respectively. Section \ref{sec: data} introduces the  datasets and empirical evidence of time-varying factor loadings. Section \ref{sec: empirical} applies the proposed methods to age-specific mortality data of  multiple countries and shows the advantages of the proposed methods. Section \ref{sec: simulation} conducts simulation studies to investigate the performance of the time-varying factor model under different scenarios. Section \ref{sec: conclusion} concludes the paper. Appendix A provides the gender-specific empirical results using the time-varying factor model. Appendix B presents the estimations of the optimal boundaries for multiple countries with a variety of forecasting horizons. Appendix C displays estimation results of the time-varying model with multiple factors. 

\section{Time-varying Factor Model}
\label{sec: model}
Let $m_{x, t}$ denote the central death rate for age $x$ in year $t$, where $x = 1, 2, \dots, N$ and $t = 1, 2, \dots, T$. Thus, $\{m_{x, t}\}_{x = 1, 2, \dots, N, t = 1, 2, \dots, T}$ is an $N$-dimensional time series with $T$ observations. Since mortality rates are always positive numbers, we use the log transformation to map the central death rates from $\mathbb{R}^{+}$ space to $\mathbb{R}$ space for modelling purposes. 
Assume $a_x$ is the age-specific constant, which is the averages over time of the $\ln(m_{x, t})$. Then $\ln(m_{x, t})-a_x$ can be modelled using the classical factor model as follows:
\begin{align}
\label{cm}
\ln(m_{x, t})= a_x+\bbb_x^{\top}\bbk_t + \varepsilon_{x, t},
\end{align}
where $\bbk_t$ is an $R \times 1$ vector of common factors; $\bbb_x$ is an $R \times 1$ vector of factor loadings, capturing the impact of each common factor on age $x$ (i.e. the age-related sensitivity to the mortality improvement); and $\varepsilon_{x, t}$ is the idiosyncratic error of $\ln(m_{x, t})$, which represents the component unexplained by the common factor. Here, $\bbk_t$, $\bbb_x$ and $\varepsilon_{x, t}$ are all unobservable components. Specifically, when $R = 1$, the classical factor model is equivalent to the Lee-Carter model. The single factor $k_t$ is defined as the mortality index in the Lee-Carter model, and consequently the factor loading $b_x$ represents the impact of the mortality index on the death rate of age $x$.

The classical factor model, however, is too restrictive when used to analyse the mortality data. It assumes that for each age the factor loadings are invariant over time. Statisticians and economists have noticed that the relationship between many economic variables and common factors is not time-invariant. Our empirical analysis using mortality data in Section \ref{sec: empirical} also suggests time-varying factor loadings.
Therefore, we develop a factor model to allow for factor loadings  changing smoothly overtime.

We introduce the time-varying factor model based on the work of \cite{su2017time}, where factor loadings are modelled as non-random functions of time. \cite{su2017time} provided a localized PCA method to consistently estimate the factors and time-varying factor loadings. Compared with \cite{park2009time}, the time-varying factor model proposed by \cite{su2017time} can capture more types of structural changes in factor loadings, including both continuous changes and abrupt structural breaks.  Assume $\ln(m_{x, t})-a_x$ follows the time-varying factor model with $R$ unobservable common factors:
\begin{align}
\label{tvm}
\ln(m_{x, t}) =a_x+ \bbb_{x,t}^{\top}\bbk_t + \varepsilon_{x, t},
\end{align}
where notations above are the same as the classical factor model, except for the factor loadings. Here, each component of the factor loading $\bbb_{x,t}$ is assumed to be a deterministic function of $t/T$: $\bbb_{x,t} = \bbb_x(t/T)$, where each component of $\bbb_x(\cdot)$ is an unknown piece-wise smooth function of $t/T$. The time-varying factor model can be seen as a generalization of the classical factor model. If $\bbb_x(\cdot)$ is invariant over time, which is a special case of the piece-wise smooth function, the time-varying factor model will degenerate to the classical factor model. Generally speaking, the assumption that factor loadings are time-invariant is too restrictive to hold in most settings. However, the time-varying factor model can relax this assumption by allowing for both continuous structural changes and abrupt changes in factor loadings, which can also benefit the mortality forecasting.

\subsection{Identification Problem}
\label{subsec: identification}
Similar to the classical factor model, there exists an identification problem in the time-varying factor model. At each time point $t$, and for any  $R \times R$ invertible matrix $\bbH_t$, we have $\bbb_{x, t}^{\top}\bbk_t = \left(\bbH_t^{-1}\bbb_{x, t}\right)^{\top}\left(\bbH_t^{\top}\bbk_t\right)$. Since an  $R \times R$ invertible matrix has $R^2$ free elements, $R^2$ restrictions are needed in parameter estimations so that $\bbb_{x, t}$ and $\bbk_t$ can be identified separately. Define $\bbB_t = \left(\bbb_{1, t}, \bbb_{2, t}, \dots, \bbb_{N, t}\right)^{\top}$ and $\bbK = \left(\bbk_{1}, \bbk_{2}, \dots, \bbk_{T}\right)^{\top}$. Then the two sets of restrictions to solve the issue of identification are  $\bbK^{\top}\bbK/T = \mathbb{I}_R$ and $\bbB_t^{\top}\bbB_t = \text{a diagonal matrix}$, where $\mathbb{I}_R$ is an $R\times R$ identity matrix. The first normalization condition imposes $R \times (R + 1)/2$ restrictions on the parameters, and the remaining $R \times (R - 1)/2$ restrictions are obtained by requiring the second constraint. These restrictions can uniquely determine the factors $\bbK$ and the factor loadings $\bbB_t$ (only up to a sign change, i.e., $-\bbK$ and $-\bbB_t$ also satisfy the two sets of restrictions). When $R = 1$, only one restriction is needed to identify parameters. We choose to use the same normalization condition as \cite{lee1992modeling}, that is, we normalize the $\bbb_{x, t}$ to sum to unity for each $t$. In this way, we can directly compare the results of our new method with that of the Lee-Carter model.

\subsection{Estimation Method}
\label{subsec: estimation}
The estimation method for the time-varying factor model is proposed by \cite{su2017time}. Let $r \in \{1, \dots, T\}$ be a fixed year. Since we have assumed that each component of $\bbb_{x, t}:[0,1] \rightarrow \mathbb{R}$ is a piece-wise smooth function, we have:
\begin{align*}
\bbb_{x, t} = \bbb_x(\frac{t}{T}) \approx \bbb_x(\frac{r}{T}) = \bbb_{x, r}, \ \text{when} \ \frac{t}{T} \approx \frac{r}{T}.
\end{align*}
Thus, the mortality rate $\ln(m_{x, t})$ can be approximated by:
\begin{align*}
\ln(m_{x, t}) \approx a_x+\bbb_{x, r}^{\top}\bbk_t + \varepsilon_{x, t}, \ \text{when} \ \frac{t}{T} \approx \frac{r}{T}.
\end{align*}
In order to estimate the factors and time-varying factor loadings, we consider the following local weighted least squares problem:
\begin{align}
\min_{\{\bbb_{x, r}\}_{x =1}^N, \{\bbk_t\}_{t=1}^T} (NT)^{-1} \sum_{x=1}^N\sum_{t=1}^T \left(\ln(m_{x, t}) -a_x- \bbb_{x, r}^{\top}\bbk_t\right)^2 K_h\left(\frac{t-r}{T}\right),
\label{eq: optim}
\end{align}
subject to the identification constraints as discussed in Section \ref{subsec: identification}. In the objective function in Equation (\ref{eq: optim}), $K_h(x) = h^{-1} K(x/h)$, where $K\left(\cdot\right)$ is a kernel function and $h$ is a smoothing parameter called ``bandwidth''. We will show that the optimization problem of Equation (\ref{eq: optim}) can be solved using the same estimation method for the classical factor model.

We have known that the mortality rates can be approximated by $\ln(m_{x, t})-a_x \approx  \bbb_{x, r}^{\top}\bbk_t + \varepsilon_{x, t}$ when $\frac{t}{T} \approx \frac{r}{T}$. Multiplying both sides of the equation by 
\begin{align*}
K_{h, tr}^{1/2} \coloneqq \left(K_h\left(\frac{t-r}{T}\right)\right)^{1/2} = \left(\frac{1}{h}K\left(\frac{t-r}{Th}\right)\right)^{1/2},
\end{align*} 
we obtain a transformed model as:
\begin{align*}
K_{h, tr}^{1/2}(\ln(m_{x, t})-a_x) \approx K_{h, tr}^{1/2}\bbb_{x, r}^{\top}\bbk_t + K_{h, tr}^{1/2}\varepsilon_{x, t}, \ \text{when} \ \frac{t}{T} \approx \frac{r}{T}.
\end{align*}
Then we can define matrices 
\begin{align*}
\bbM^{(r)} = \left(\bbM^{(r)}_1, \dots, \bbM^{(r)}_N\right),
\end{align*}
\begin{align*}
\bvar^{(r)} = \left(\bvar_{1}^{(r)}, \dots, \bvar_{N}^{(r)}\right),
\end{align*}
and
\begin{align*}
\bbK^{(r)} = \left(K_{h, 1r}^{1/2}\bbk_1, \dots, K_{h, Tr}^{1/2}\bbk_T\right)^{\top}.
\end{align*}
where $\bbM^{(r)}_x = \left(K_{h, 1r}^{1/2}(\ln(m_{x, 1})-a_x), \dots, K_{h, Tr}^{1/2}(\ln(m_{x, T})-a_x)\right)^{\top}$ and $\bvar_{x}^{(r)} = \left(K_{h, 1r}^{1/2}\varepsilon_{x, 1}, \dots, K_{h, Tr}^{1/2}\varepsilon_{x, T}\right)^{\top}$ with $x=1, 2, \ldots, N$.
Therefore, the transformed model can be written in matrix form as follows:
\begin{align*}
\bbM^{(r)} \approx \bbK^{(r)}\bbB_r^{\top} + \bvar^{(r)},
\end{align*}
and the optimization problem above can also be written in matrix notation as:
\begin{align*}
&\min_{\bbK^{(r)},\bbB_r} \Tr{\left(\left(\bbM^{(r)} - \bbK^{(r)}\bbB_r^{\top}\right)\left(\bbM^{(r)} - \bbK^{(r)}\bbB_r^{\top}\right)^{\top}\right)}\\
& s.t \ \bbK^{(r)\top}\bbK^{(r)}/T = \mathbb{I}_R \ \text{and} \ \bbB_r^{\top}\bbB_r = \text{a diagonal matrix}.
\end{align*}
Concentrating out $\bbB_r = \bbM^{(r)\top} \bbK^{(r)}\left(\bbK^{(r)\top}\bbK^{(r)}\right)^{-1}$ (which is $\bbM^{(r)\top} \bbK^{(r)}/T$ under the normalization
$\bbK^{(r)\top}\bbK^{(r)}/T = \mathbb{I}_R$), the optimization problem is converted to minimizing the objective function:
\begin{align*}
\Tr{\left(\bbM^{(r)\top}\bbM^{(r)}\right)} - T^{-1} \Tr{\left(\bbK^{(r)\top}\bbM^{(r)}\bbM^{(r)\top} \bbK^{(r)}\right)}.
\end{align*}
Thus, the original local weighted least squares problem is equivalent to maximizing 
\begin{align*}
\Tr{\left(\bbK^{(r)\top}\bbM^{(r)}\bbM^{(r)\top} \bbK^{(r)}\right)},
\end{align*}
subject to the restriction $\bbK^{(r)\top}\bbK^{(r)}/T = \mathbb{I}_R$, which is equivalent to the optimization problem of the classical factor model. 

Our objective is to obtain estimators of the factors and factor loadings. A two-stage estimation procedure is used to estimate those parameters. Let $\widehat{\bbK}^{(r)}$ denote the estimated factor matrix of $\bbK^{(r)}$, and $\widehat{\bbB}_r = \left(\widehat{\bbb}_{1,r}, \widehat{\bbb}_{2,r}, \dots, \widehat{\bbb}_{N,r}\right)^{\top}$ denote the estimator of the time-varying factor loading matrix $\bbB_r$. Then, $\widehat{\bbK}^{(r)}$ is $\sqrt{T}$ times eigenvectors corresponding to the largest $R$ eigenvalues of the $T \times T$ matrix $\bbM^{(r)}\bbM^{(r)\top}$, and $\widehat{\bbB}_r$ is $\bbM^{(r)\top}\widehat{\bbK}^{(r)}\left(\widehat{\bbK}^{(r)\top}\widehat{\bbK}^{(r)}\right)^{-1}$ (it is $\bbM^{(r)\top}\widehat{\bbK}^{(r)}/T$ under the condition $\bbK^{(r)\top}\bbK^{(r)}/T = \mathbb{I}_R$). Therefore, in the first step, we can acquire estimators $\widehat{\bbB}_r$ of the factor loadings for $r = 1, \dots, T$.

  Based on the estimator $\widehat{\bbB}_r$ of the factor loading matrix obtained in the first stage, we consider another least squares problem in the second stage to obtain the estimator of the factor $\bbk_t$. The objective function we would like to minimize is as follows:
\begin{align*}
\sum_{x = 1}^N \left(\ln(m_{x, t}) -a_x- \widehat{\bbb}_{x,t}^{\top}\bbk_t\right)^2 \ \text{for} \ t = 1, \dots, T.
\end{align*}
Since we already have $\widehat{\bbb}_{x,t}$ in the first stage, the answer to this minimization problem is
\begin{align*}
\widehat{\bbk}_t = \left(\sum_{x = 1}^N \widehat{\bbb}_{x, t} \widehat{\bbb}_{x, t}^{\top}\right)^{-1} \left(\sum_{x = 1}^N \widehat{\bbb}_{x, t}\left(\ln(m_{x, t})-a_x\right)\right) \ \text{for} \ t = 1, \dots, T.
\end{align*}
Thus, using the two-stage estimation method, we can obtain consistent estimators for both the factors and time-varying factor loadings. 

Next, we discuss some issues in the kernel estimation. 
\begin{rmk}
\textbf{Boundary kernel}. Usually, there exists a boundary bias issue in the kernel estimation. Instead of using the ordinary kernel function, it is suggested that a boundary kernel should be used to help us obtain some uniform results. Let $\lfloor a \rfloor$ represent the greatest integer less than or equal to $a$, then the boundary kernel we choose to use is as follows:
\begin{align*}
K_{h,tr}^* = h^{-1}K_r^*\left(\frac{t-r}{Th}\right) = \begin{cases}
    \frac{h^{-1}K\left(\frac{t-r}{Th}\right)}{\int_{-\frac{r}{Th}}^1 K(u) \mathrm{d}u}       & \quad r \in [1, \lfloor Th \rfloor]\\
    h^{-1}K\left(\frac{t-r}{Th}\right),  & \quad r \in (\lfloor Th \rfloor, T - \lfloor Th \rfloor] \\
    \frac{h^{-1}K\left(\frac{t-r}{Th}\right)}{\int_{-1}^{(1-r/T)/h} K(u) \mathrm{d}u} & \quad r \in (T - \lfloor Th \rfloor, T]
  \end{cases}.
\end{align*}
\end{rmk}

\begin{rmk}
\textbf{The choice of bandwidth}. For the nonparametric local smoothing method, it is important to determine the bandwidth for the kernel estimation. There are two ways to choose the bandwidth. The first one is to use a data-driven method, such as the cross-validation. The other one is to use Silverman’s rule of thumb to set the bandwidth, which is much easier to compute. \cite{su2017time} have shown that choices of the kernel function and the bandwidth have little impact on the performance of the information criteria. Thus, in the following empirical analysis, we decide to use the Epanechnikov kernel and its corresponding Silverman’s rule of thumb bandwidth, which is $h = (2.35/\sqrt{12})T^{-1/5}N^{-1/10}$.
\end{rmk}

\begin{rmk}
\textbf{Determination of the number of factors}.
There are mainly two methods to determine the number of factors, $R$. The first one is to use a BIC-type information criterion proposed by \cite{su2017time}. Under certain assumptions, the new information criterion can correctly choose the true value of $R$. However, those assumptions may not hold in real data. Additionally, it is not easy to implement the out-of-sample forecasting if the chosen value of $R$ is too large. 

The second method is based on the fact that the original local weighted least squares problem can be transformed into an optimization problem of the classical factor model. Therefore, the cumulative sum of eigenvalues can help us  identify the number of factors. Let $c$ denote a cut-off value between $0$ and $1$, and $\lambda_k$ denote the $k^{th}$ largest eigenvalue of the matrix $\bbM^{(r)}\bbM^{(r)\top}$, then we can choose the value of $R$ as $\min\{R: \left(\sum_{k=1}^R \lambda_k\right)/\left(\sum_{k=1}^{N} \lambda_k\right) \ge c\}$. In the following analysis, we will set the cut-off value as $c=0.9$ and empirical analysis shows that only one factor is enough to capture most characteristics of the mortality data, which is consistent with the \cite{lee1992modeling} model.
\end{rmk}

\subsection{Forecasting Method}
\label{subsec: forecast}
We now consider how to make out-of-sample forecasting using the time-varying factor model. Since the factor loadings change over time, we should not only make predictions of the common factors, but also extrapolate the factor loadings for each age.
We describe the forecasting method for a single factor model ($R = 1$) for the simplicity of notations  in the following analysis.
Assume that based on the historical data we have acquired the estimated common factor and factor loadings using the method mentioned in Section \ref{subsec: estimation}. 

In order to forecast the common factor, we first fit the common factor with an   ARIMA model. Since Akaike Information Criterion (AIC) is asymptotically equivalent to the cross-validation when the maximum likelihood estimation is used to fit the model \citep{stone1977asymptotic}, we choose AIC as the model selection criteria to find the most appropriate ARIMA model. After that, we can use the chosen model to forecast and obtain prediction intervals for the latent factor (see more details in Chapter 5 \& 9 of \cite{brockwell1991time}).

The factor loading $b_{x,t}$ is assumed to be an unknown piece-wise smooth function of time $t$. 
For the purpose of extrapolating the factor loading $b_{x,t}$ into the future, we will adopt two different methods to achieve the goal:

\begin{enumerate}
\item \textit{The naive method}. We simply assume that in the forecasting horizon, $b_{x,t} (t > T)$ is set as $b_{x,T}$. This is essentially a parametric forecasting method and similar to that in the Lee-Carter model but with a different estimated value. The naive method using constant factor loading has a simple structure and could provide more stable forecasts in the long term.
\item \textit{The local regression method}. This method is based on a nonparametric regression method -- the local linear regression, to flexibly estimate the deterministic function $b_{x,t}$ (See more details in \citet{fan1996local} and \citet{friedman2001elements}). Similar method has also been applied in \citet{LI2017166, LI2015264}. The local linear regression can easily extend the most recent trends, which is more suitable for short-term forecasting.
\end{enumerate}

We briefly describe the local regression method in the rest of this section. The main idea of the local linear regression is to fit the linear regression using only the observations in the neighbourhood of a target point $t_0$. This so-called localization is achieved by using a weight function $K_{\lambda}(t, t_0) = K\left((t-t_0)/\lambda\right)$, where $K$ is a  kernel function and the index $\lambda$ indicates the width of the neighborhood. One of the commonly used kernel functions with compact support is Epanechnikov kernel, which is adopted in this paper. For the Epanechnikov kernel, the window width parameter $\lambda$ is the radius of the support region, which can be estimated using out-of-sample validation. The weight function assigns a weight to each time point $t$ based on the corresponding distance from $t_0$ (i.e., $|t- t_0|$). In this way, the resulting estimated function is a smooth function.

Specifically for the forecasting of the time-varying factor loading of each age $x$, the local linear regression solves a separate weighted least square problem at each target point $T+h$ ($h = 1, 2, \dots)$:
\begin{align*}
\min_{\alpha(T+h), \beta(T+h)} \sum_{t = 1}^{T+h-1} K_{\lambda}(t, T+h) \left(b_{x, t} - \alpha(T+h) - \beta(T+h)t\right)^2.
\end{align*}
Note that the notations $\alpha(T+h)$ and $\beta(T+h)$ indicate that the two parameters under study vary with the point $T+h$ in the local linear method. 

Let $\bbb_x = \left(b_{x,1}, b_{x,2}, \dots, b_{x,T + h -1}\right)^{\top}$, $\bbX = \begin{pmatrix}
	1&1&\dots&1 \\
	1&2&\dots&T + h -1
\end{pmatrix}^{\top}$, and $\bbW(T+h)$ denote the $(T+h-1) \times (T+h-1)$ diagonal matrix with the $t^{th}$ diagonal element $K_{\lambda}(t, T+h)$. Then by using the weighted least squares estimation, we can obtain the estimators for $\alpha(T+h)$ and $\beta(T+h)$ as follows:
\begin{align*}
\left(\widehat{\alpha}(T+h), \widehat{\beta}(T+h) \right)^{\top} = \left(\bbX^{\top}\bbW(T+h)\bbX\right)^{-1} \bbX^{\top}\bbW(T+h)\bbb_x.
\end{align*}
To ensure that $\bbX^{\top}\bbW(T+h)\bbX$ is nonsingular, the bandwidth parameter $\lambda$ in the kernel function should be selected properly in practice, see more details in \citet{fan1996local}.
Therefore the forecasted factor loading at point $T+h$ is
\begin{align*}
\widehat{b}_{x, T+h} = \begin{pmatrix} 1 & T+h \end{pmatrix} \begin{pmatrix} \widehat{\alpha}(T+h)\\ \widehat{\beta}(T+h) \end{pmatrix}.
\end{align*}
Note that for $h > 1$, the forecasts $\widehat{b}_{x, T+1}, \dots, \widehat{b}_{x, T+h-1}$ are evolved in $\bbb_x$ when estimating the factor loading at the time $T+h$. Following this method, we can estimate the factor loadings for each age as a smooth function of time $t$ and then extrapolate the factor loadings into the future. Combining with the predicted common factors, we can make out-of-sample predictions of the central death rates using the time-varying factor model.

\section{Optimal `Boundary' Estimation}
\label{sec: choice}



Under the framework of time-varying factor model, we assume the factor loading $b_{x,t}$ is a function of time $t$. In Section \ref{subsec: forecast}, we introduced two different methods to extrapolate the factor loading. One is a naive method, which is more suitable for long-term forecasting; and the other is based on local linear regression, which is more suitable for short-term forecasting. Then can we estimate the `boundary' between short-term and long-term forecasting that divides the forecasting horizon according to the predictive power of the local regression method and naive method?

To solve this problem, we first propose a new forecasting method, which is a hybrid of two previously introduced methods. Assume $T_0$ is the number of years used in fitting the model and $k$ ($k =  0,1,2,\cdots$) is the optimal boundary between short-term and long-term forecasting, favoured by the time-varying models based on the local regression and naive method, respectively. We have the point forecast estimation of mortality rate $ln(m_{x, t})$ for any given $x$, $t$ and $k \ (k \ge 1)$ using the hybrid method  as 
\begin{equation*}
ln(\widehat{m}_{x, t})=\begin{cases}
\widehat{a}_{x}+\widehat{b}_{x, t} \cdot \widehat{k}_{t} & T_0 + 1 \le t \le T_0+k, \\
\widehat{a}_{x}+\widehat{b}_{x, T_0+k} \cdot \widehat{k}_{t} & t \ge T_0+k+1.
\end{cases}
\end{equation*}
If $T_0+1 \le t \le T_0+k$, the forecast of $\ln(m_{x, t})$ at time $t$ is $\widehat{a}_{x}+\widehat{b}_{x, t} \cdot \widehat{k}_{t}$, where $\widehat{b}_{x, t}$ is the extrapolated factor loading at time $t$ based on the local regression method. When $t \ge T_0+k+1 $, the forecast at time $t$ is $ \widehat{a}_{x}+\widehat{b}_{x, T_0+k} \cdot \widehat{k}_{t}$, where $\widehat{b}_{x, T_0+k}$ is time-invariant and obtained using the extrapolated factor loading at time $T_0+k$ based on the local regression method. 
For $k = 0$, the forecast at time $t \ (t>T_0)$ is just $\widehat{a}_{x}+\widehat{b}_{x, T_0} \cdot \widehat{k}_{t}$ using the estimated factor loading at time $T_0$. In this case, the hybrid method degenerates to the naive method. 
In view of this, $T_0+k$ is  the time boundary between short-term and long-term forecasting, and between choosing the local regression and naive method. Given the value of  $k$, the hybrid method applies the local linear regression for the first $k$ periods in the forecasting horizon and keeps the factor loadings ($\widehat{b}_{x, T_0+k}$) unchanged thereafter, which combines the local regression and naive methods. Additionally, the hybrid method guarantees a consistent and smooth transition from short-term to long-term forecasting.

 As discussed in Section \ref{sec: intro}, different forecasting horizons may favour different models. Generally, long-term forecasting  benefits more from the historical long-term trend and short-term forecasting relies on the recent trend \citep{booth2002applying}. Since the local linear regression can easily extend the most recent trend, it is more suitable for short-term forecasting. However, as time goes by, the recent trend becomes less and less reliable, which is not suitable for long-term forecasting. On the other hand, the naive method using constant factor loading is more suitable for long-term forecasting, as it has a simple structure and  would provide more stable forecasts in the long term. Compared to the classical factor model, the naive method provides more accurate estimations not only for the factor loadings but also for the common factors, which helps generate more accurate long-term forecasts.

Based on the hybrid forecasting method, we  propose an estimation method of the optimal `boundary' inspired by  \cite{bai2010common}. Assume the entire dataset has $T$ years and we consider the first $T_0$ years of  data as the training set, and  the remaining data with size $T-T_0$ as the validation set. Given the value of $k$, we first fit the time-varying factor model using the training set, and then apply the hybrid forecasting method to the validation set. We consider all possible lengths of short-term (long-term) forecasting horizons (i.e. $k$) and find out an optimal one using least squares estimation. We describe the estimation procedure as follows.

For the given $x$ and $k$ such that $1 \le k \le T-T_0-1$, define $\widehat{y}_{x, t}(k) = \widehat{a}_{x}+\widehat{b}_{x, t} \cdot \widehat{k}_{t}$ as the predicted value of $\ln(m_{x, t})$ using the hybrid forecasting method. When $T_0+1 \le t \le T_0+k$, $\widehat{b}_{x, t}$ is forecasted by the local regression method; And when $T_0+k+1 \le t \le T$, $\widehat{b}_{x, t} = \widehat{b}_{x, T_0+k}$, where $\widehat{b}_{x, T_0+k}$ is the predicted factor loading at time $T_0+k$ obtained via the local regression method. Then we define the sum of squared residuals for age $x$ as
\begin{align*} 
S_{x, T}(k)&=\sum_{t=T_0+1}^{T}\left(\ln(m_{x, t})-\widehat{y}_{x, t}(k)\right)^2 \\
& =  \sum_{t=T_0+1}^{T_0+k}\left(\ln(m_{x, t})-\widehat{a}_{x}-\widehat{b}_{x, t} \cdot \widehat{k}_{t}\right)^2 +  \sum_{t=T_0+k+1}^{T}\left(\ln(m_{x, t})-\widehat{a}_{x}-\widehat{b}_{x, T_0+k} \cdot \widehat{k}_{t}\right)^2,
\end{align*}
where $k=1, 2, \dots, T-T_0-1$. Here $k$ represents the length of the short-term forecasting horizon or the `boundary' between short-term (based on the local regression method) and long-term (based on the naive method) forecasting. The local linear regression is used to make forecasts from  $T_0+1$ to $T_0+k$; while the naive method (i.e. assuming $\widehat{b}_{x, t}$ does not change over the remaining period) is used to make forecasts from $T_0+k+1$ to  $T$. We define
\begin{align*} 
S_{x, T}(0)=\sum_{t=T_0+1}^{T}\left(\ln(m_{x, t})-\widehat{a}_{x}-\widehat{b}_{x, T_0} \cdot \widehat{k}_{t}\right)^2, \ \ \text{for} \ \  k = 0 
\end{align*}
and
\begin{align*}
S_{x, T}(T-T_0)= \sum_{t=T_0+1}^{T}\left(\ln(m_{x, t})-\widehat{a}_{x}-\widehat{b}_{x, t} \cdot \widehat{k}_{t}\right)^2, \ \ \text{for} \ \  k= T-T_0.
\end{align*}
In this way, $S_{x, T}(k)$ is defined for each $k = 0,1, \dots, T-T_0$. Thus, the total sum of squared residuals (SSR) across all ages is defined as 
\begin{align*}
SSR(k)=\sum_{x=1}^{N} S_{x,T}(k),
\end{align*} 
Hence the least squares estimator of the optimal `boundary' is
\begin{align*}
\widehat{k}= \underset{0\le k \le T-T_0}{argmin}SSR(k).
\end{align*}
The estimated optimal `boundary' between the short-term (based on local linear regression) and long-term (based on naive method) forecasting is the time $\widehat{k}$ that leads to the smallest SSR.

\section{Data}
\label{sec: data}
The mortality data used in this paper are extracted from the Human Mortality Database (HMD) (\citenum{HMD}). Six countries are selected for the empirical analysis in Section \ref{sec: data} and Section \ref{sec: empirical}. For each country, age-sex-specific death rates are available annually for the entire population. The selected countries are shown in Table \ref{t1} along with the corresponding available time horizons, which will be used for empirical analysis.

\begin{table}[!htbp] \centering 
	\small
	\caption{Time horizon for different countries} 
	\label{t1} 
	\begin{tabularx}{\textwidth}{c *{4}{Y}} 
		\toprule
		Country & start year & end year & length \\
		\midrule 
		AUSTRALIA & 1921       & 2018    & 98  \\
		CANADA & 1921       & 2016    & 96  \\
		FRANCE & 1816       & 2017    & 202  \\
		ITALY & 1872       & 2017    & 146  \\
		JAPAN & 1947       & 2018    & 72  \\
		USA & 1933      & 2017    & 85  \\
		\bottomrule 
	\end{tabularx} 
\end{table} 

The mortality data are generally available from age $0$ to age $110+$ for each year. Since measures of mortality at very old ages are unreliable \citep{lee1992modeling}, we decide not to use mortality data of age $91$ and over in the following analysis and end up with $N = 91$ ages.   

\begin{figure}[h]
	\centering
	\begin{minipage}[c]{0.3\textwidth}
		\includegraphics[width=\linewidth]{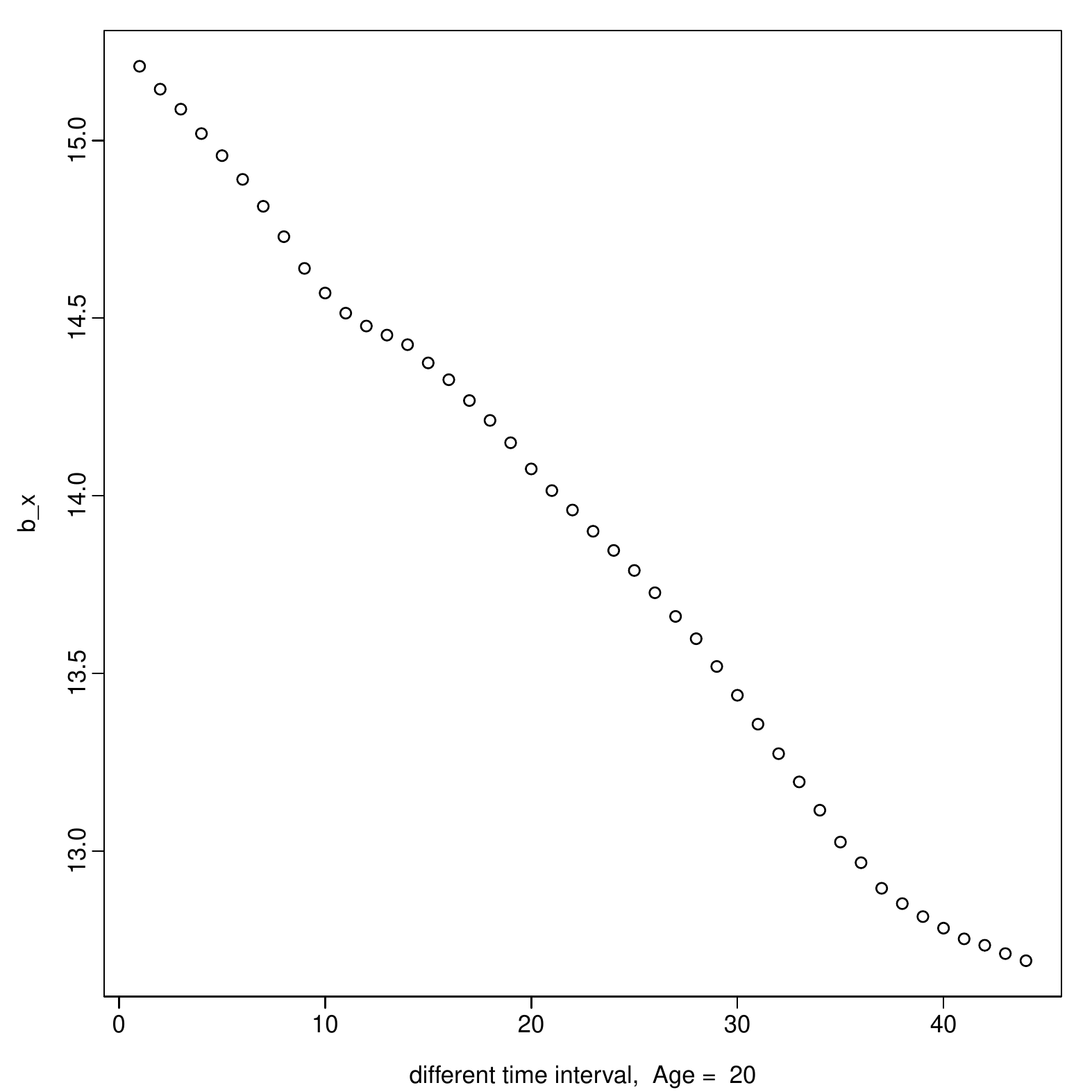}
	\end{minipage}
	\hspace*{0.08cm}
	\begin{minipage}[c]{0.3\textwidth}
		\includegraphics[width=\linewidth]{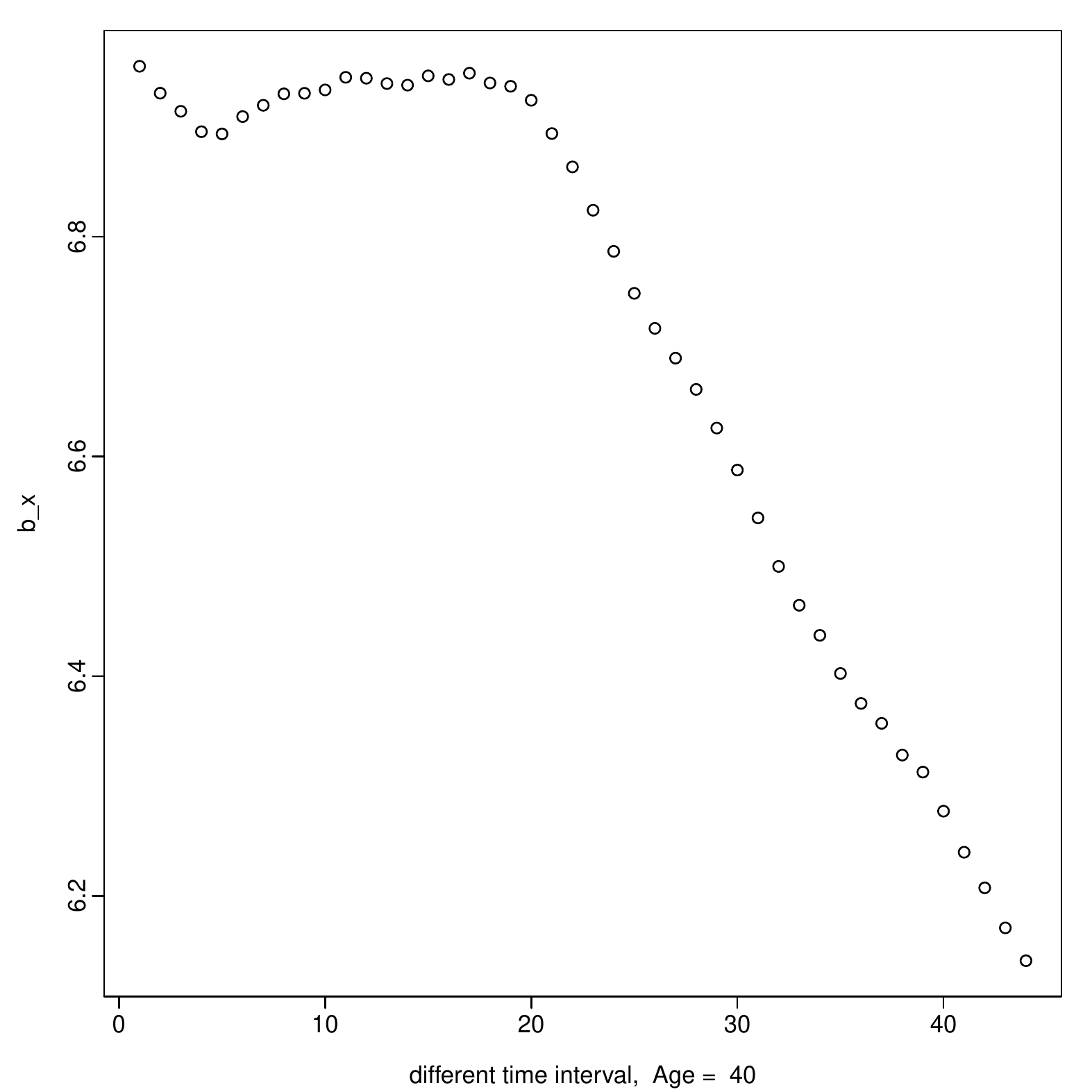}
	\end{minipage}
	\\
	\begin{minipage}[c]{0.3\textwidth}
		\includegraphics[width=\linewidth]{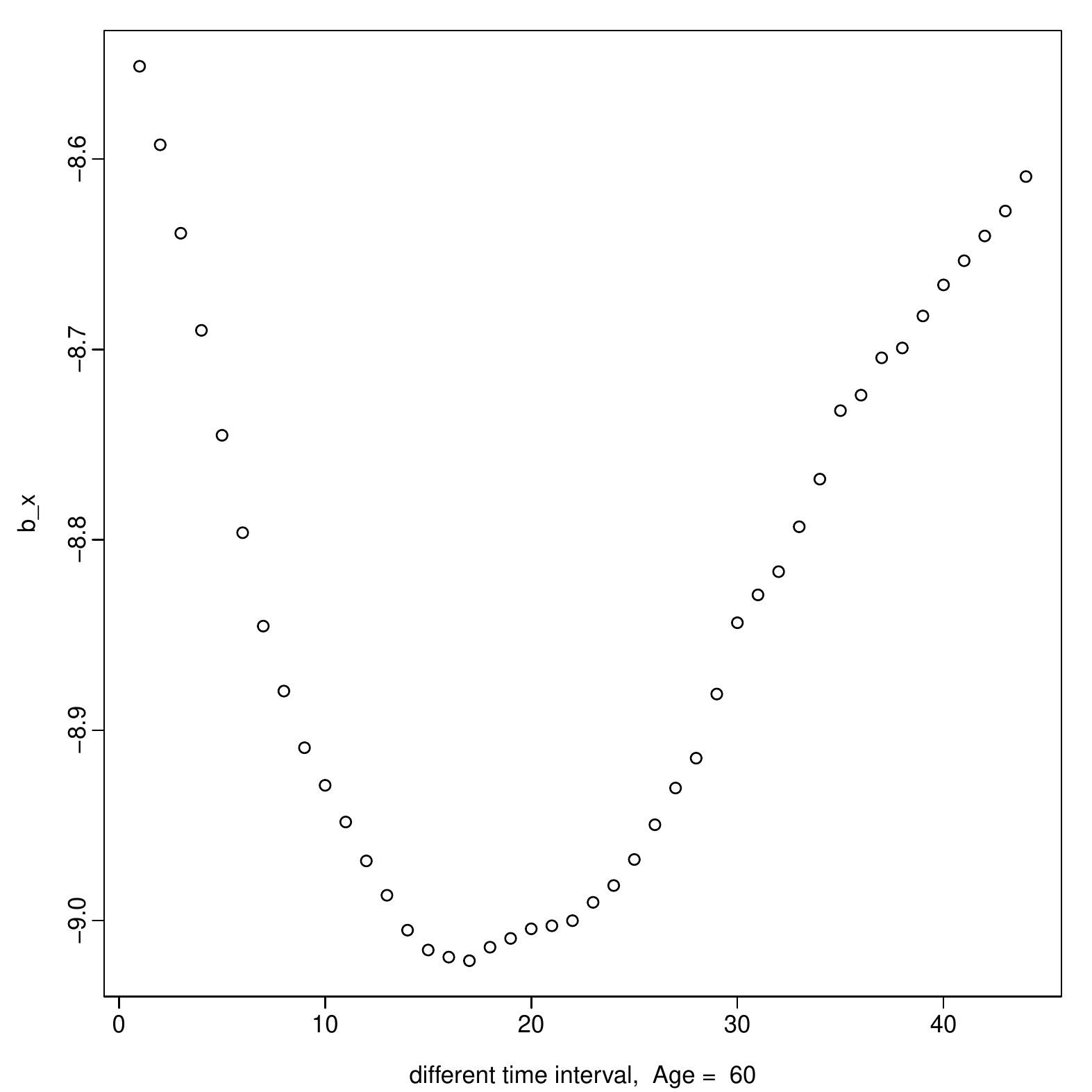}
	\end{minipage}
	\hspace*{0.08cm}
	\begin{minipage}[c]{0.3\textwidth}
		\includegraphics[width=\linewidth]{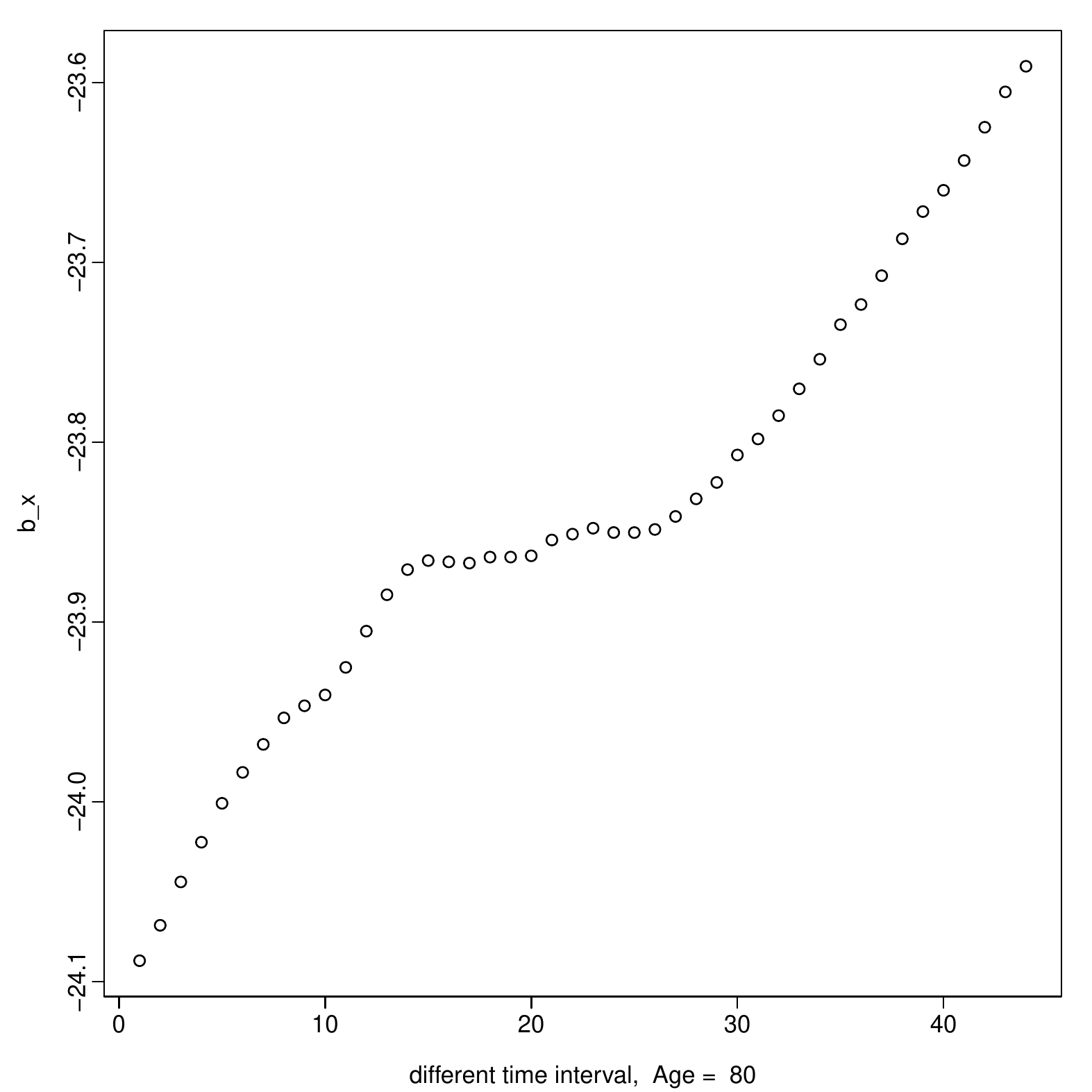}
	\end{minipage}
	\caption{Factor loadings for ages 20, 40, 60, 80 over 44 rolling-window time frames}
	\label{bx}
\end{figure}

In order to investigate whether the factor loadings are time-varying or time-invariant in the empirical data. We conduct an exploratory data analysis by applying the Lee-Carter Model on the US mortality data with rolling-window time frames. We first divide the entire dataset into 44 subsets (each with 40 yearly observations) with the first subset from year 1933 to year 1972, the second subset from year 1934 to year 1973, and so on. We then fit the Lee-Carter model on each of the subset and extract the factor loading $b_x$ for each time frame. We plot the factor loadings of some selected  ages  in Figure \ref{bx}. We can see that the factor loadings possess different dynamic patterns for different ages and they are not time-invariant.

\section{Empirical Results and Analysis}
\label{sec: empirical}
In the first two subsections, we present the application results of the time-varying factor model using age-specific mortality data of the US. We compare the time-varying factor models based on both the naive and local regression forecasting methods with Lee-Carter model (the classical factor model with one factor) via out-of-sample forecasting performance. Empirical results by gender are provided in Appendix A. Section \ref{subsec: comparison} further compares the forecasting performance across multiple countries and models based on different fitting and forecasting horizons. And in the Section \ref{subsec: change}, we estimate the optimal `boundary' between short-term and long-term forecasting for different countries.

\subsection{Model Fitting}
\label{subsec: fit}

We fit the US mortality data using the estimation method of the time-varying factor model introduced in Section \ref{sec: model}. The number of factors estimated is 1 ($\widehat{R}=1$), which is consistent with the Lee-Carter model. More empirical results of the time-varying model with multiple factors are shown in Appendix C. 

Using the model selection criteria AIC, we find that the common factor $k_t$, obtained from the time-varying factor model, follows an ARIMA $(1,1,0)$ with drift model. This model can capture most of the characteristics of the common factor. Our fitted model of the common factor $k_t$ is as follows:
\begin{align*}
\bigtriangledown k_t = \underset{(0.2791)}{-1.4116} + \underset{(0.1046)}{0.3271} \bigtriangledown k_{t-1} + e_t,
\end{align*}
where $\bigtriangledown$ refers to the first order differencing and $e_t$ represents the error term. The numbers in the parentheses are the standard errors of the corresponding parameters. With the ARIMA model built above, we can then forecast the common factor into the future.

\begin{figure} 
	\centering
	\includegraphics[width=0.5\linewidth]{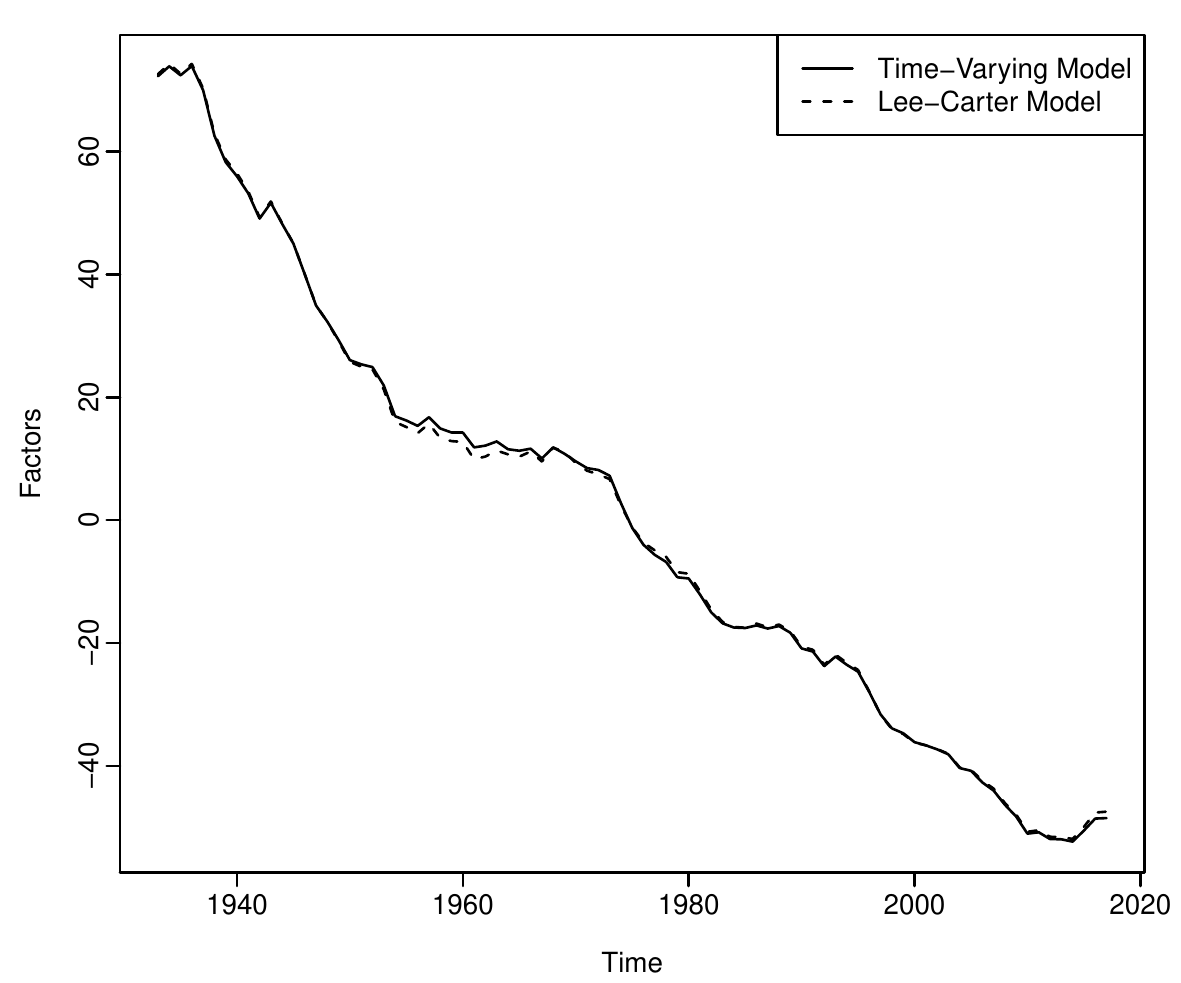}
	\caption{Plots of the estimated common factors for the time-varying factor model \& the Lee-Carter model}
	\label{f2}
\end{figure}

As a comparison, we also list the fitted ARIMA model of the common factor (The number of factors estimated is 1.) using the classical factor model\footnote{In this case, the classical factor model has the same model structure as the Lee-Carter model with $R=1$.}  below:
\begin{align*}
\bigtriangledown k_t = \underset{(0.2837)}{-1.4046} + \underset{(0.1051)}{0.3114} \bigtriangledown k_{t-1} + e_t.
\end{align*}
We can see that the ARIMA models of the common factors estimated from the time-varying factor model and the Lee-Carter model are close. The estimated common factors, plotted in Figure \ref{f2}, tend to decrease linearly 
 and show similar dynamic patterns. The common factor is regarded as the index of the level of mortality, which captures major influence on death rates of all ages.

\begin{figure}[hbt!]
	\begin{minipage}{0.45\textwidth}
		\includegraphics[width=\linewidth, page=1]{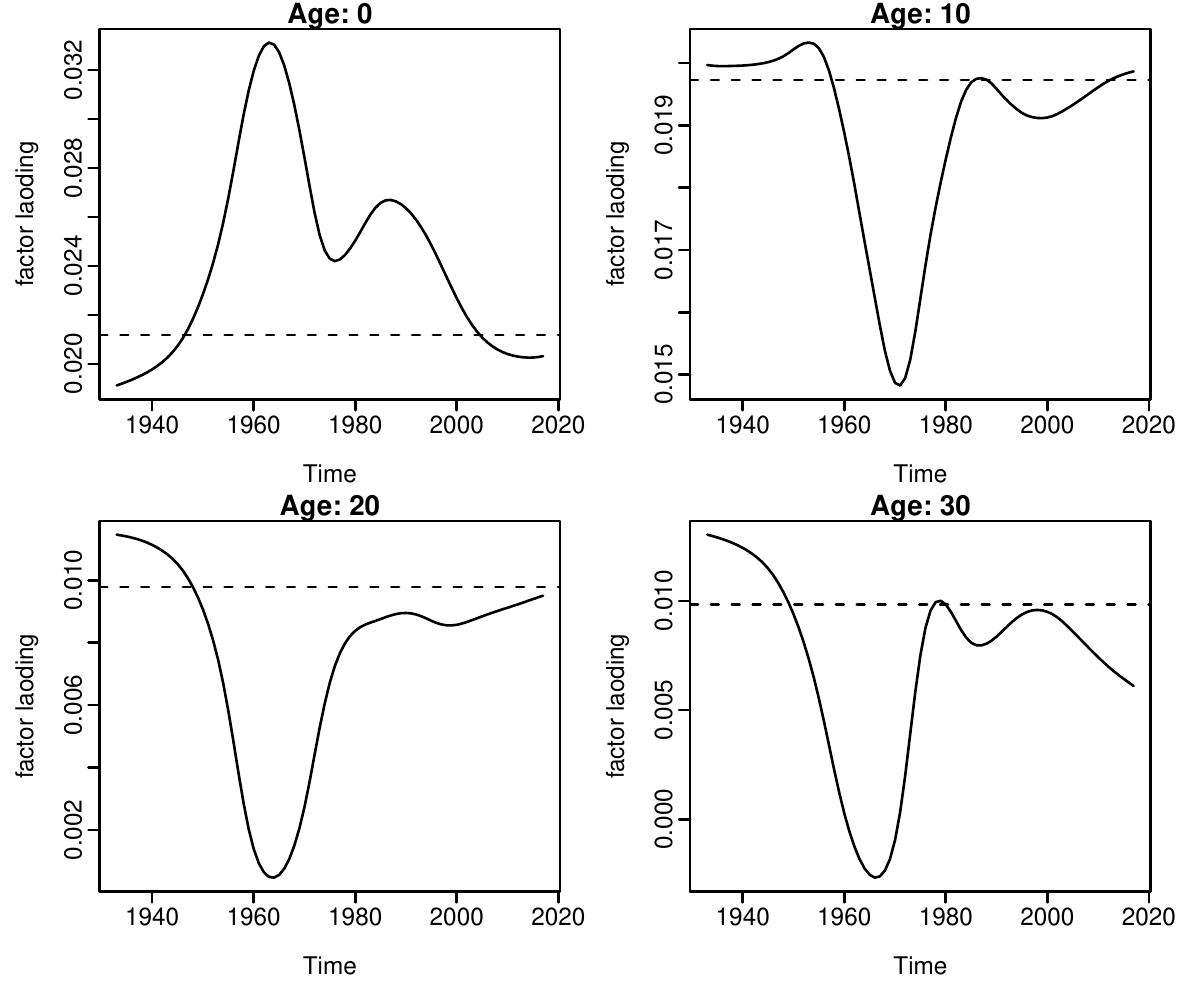}
		\includegraphics[width=\linewidth, page=2]{fig/f3_1.pdf}
		\includegraphics[width=\linewidth, page=1]{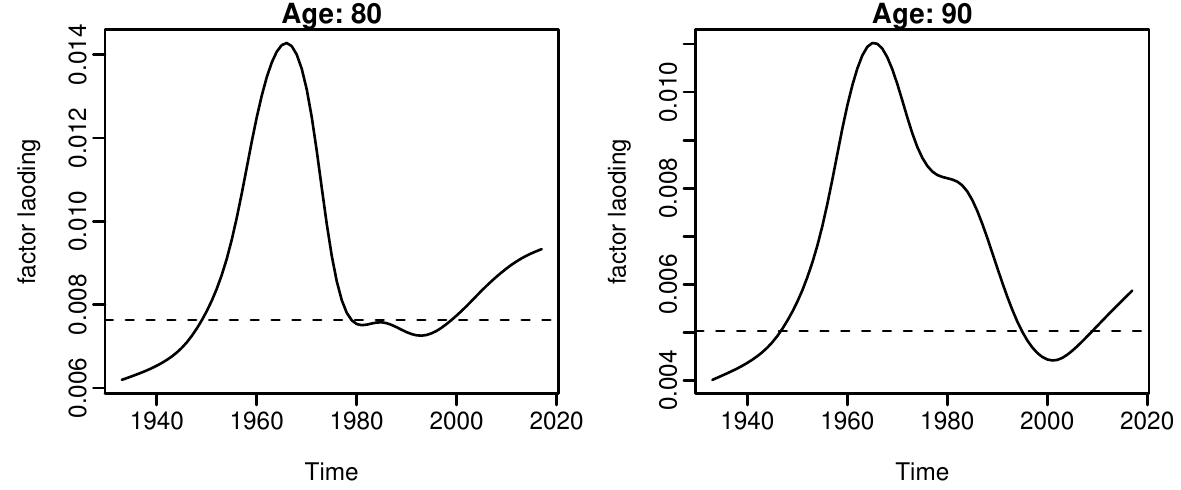}
		\caption{Plots of the estimated time-invariant factor loadings (dashed lines) \& the time-varying factor loadings (solid lines) for age $0, 10, \dots, 90$.
		}
		\label{f3}
	\end{minipage}%
	\hfill
	\begin{minipage}{0.45\textwidth}
		\includegraphics[width=\linewidth, page=1]{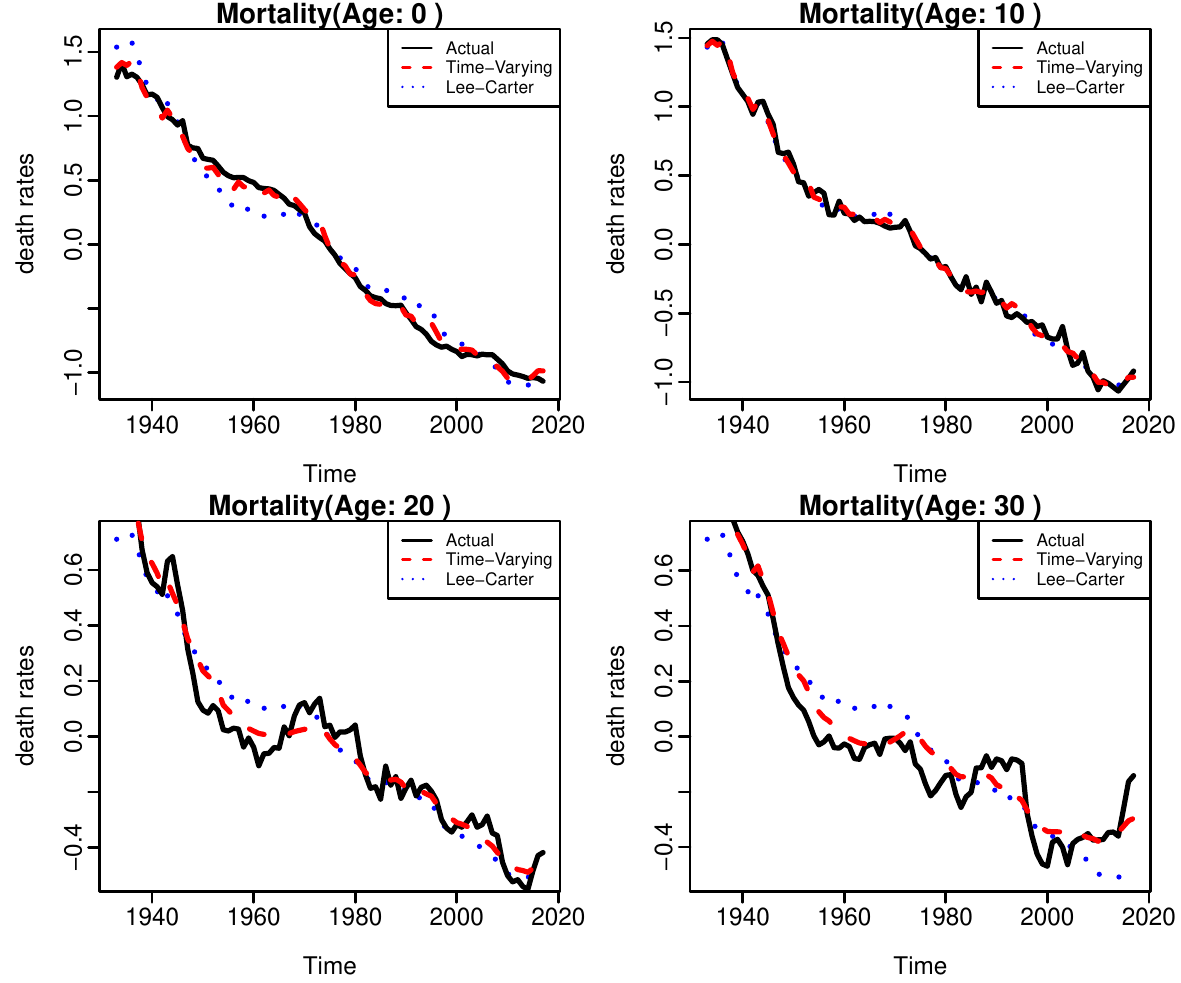}
		\includegraphics[width=\linewidth, page=2]{fig/f4_1.pdf}
		\includegraphics[width=\linewidth, page=1]{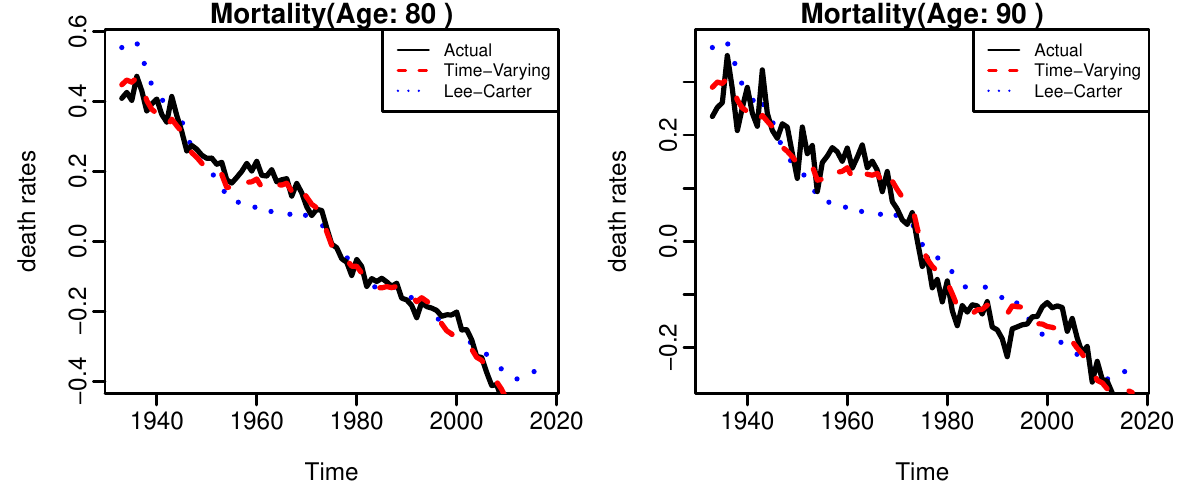}
		\caption{The actual data (black solid lines) versus the fitted values from the time-varying model (red dashed lines) and the Lee-Carter model (blue dotted lines); the data have been log-transformed \& demeaned.
		}
		\label{f4}
	\end{minipage}
\end{figure}

Figure \ref{f3} displays the comparison of the factor loadings between the time-varying factor model and the Lee-Carter model for selected ages. Compared with time-invariant factor loadings (the dashed lines), the time-varying factor loadings (the solid curves) change smoothly overtime, see Figure \ref{f3}. It is interesting to notice that, no matter which age it is, the corresponding factor loadings always reach their own minimum or maximum values during $1960$s or $1970$s. For older people (over age $40$), the factor loadings usually arrive at their  maximum values during $1960$s or $1970$s, which means the  death rates of older people are more sensitive to the latent common factor during that period.  For the younger ages (below age $40$), however, the corresponding factor loadings reach their minimum values during the same period, which means the death rates of younger people are less sensitive to the latent factor during that time. The only exception is the factor loadings of the infant group, whose dynamic pattern is more similar to that of the older group. 

 Figure \ref{f4} shows the fitted death rates of both the time-varying factor model and the Lee-Carter model with empirical observations for selected ages. Obviously, no matter which age it is, the time-varying factor model fits better than the Lee-Carter model. We use the mean squared error (MSE)\footnote{The MSE for the time-varying model is computed as follows:
 	\begin{align*}
 	\text{MSE} = \frac{1}{NT} \sum_x\sum_t\left(\ln(m_{x, t}) - a_x-\widehat{\bbb}_{x,t}^{\top}\widehat{\bbk}_t\right)^2
 	\end{align*}
 	Computation of the MSE for the Lee-Carter model is the same as above except that $\widehat{\bbb}_{x,t}$ is replaced by $\widehat{\bbb}_{x}$.} to evaluate the goodness of fit.  As a result, the overall MSE of the time-varying factor model is $0.001990$, which is much smaller than that of the Lee-Carter model, $0.006690$ (three times bigger than the former one). Therefore, the time-varying factor model performs much better than the Lee-Carter with respect to the in-sample fitting.

Although the time-varying factor model works better in the fitting procedure, the problem of overfitting may exist due to the increased complexity of the model.  Through the Monte Carlo simulation studies in Section \ref{sec: simulation}, we will see that overfitting is harmful to forecasting. Usually, an overfitting model performs too well in the fitting sample to have good generalization ability in forecasting. Generally speaking, we can always improve a model’s in-sample fitting performance by increasing the complexity of the model, which, however, cannot guarantee a better forecasting performance in the future. Thus we will use the out-of-sample validation method to investigate whether the time-varying model can enhance the out-of-sample forecasting performance in the next subsection.

\subsection{Out-of-sample Forecasting}
 In this subsection, we use the original US mortality data over the first 60 years as the training set (from 1933 to 1992) to fit the models, and then forecast the mortality rates in the testing set (from 1993 to 2017) using the fitted models. The predicted values are compared with the actual data in the testing set to see which model is better at the out-of-sample forecasting. We apply the mean squared prediction error (MSPE) as the measure to evaluate the out-of-sample forecasting performance.
 
 \begin{figure} 
	\centering
	\includegraphics[width=0.6\linewidth]{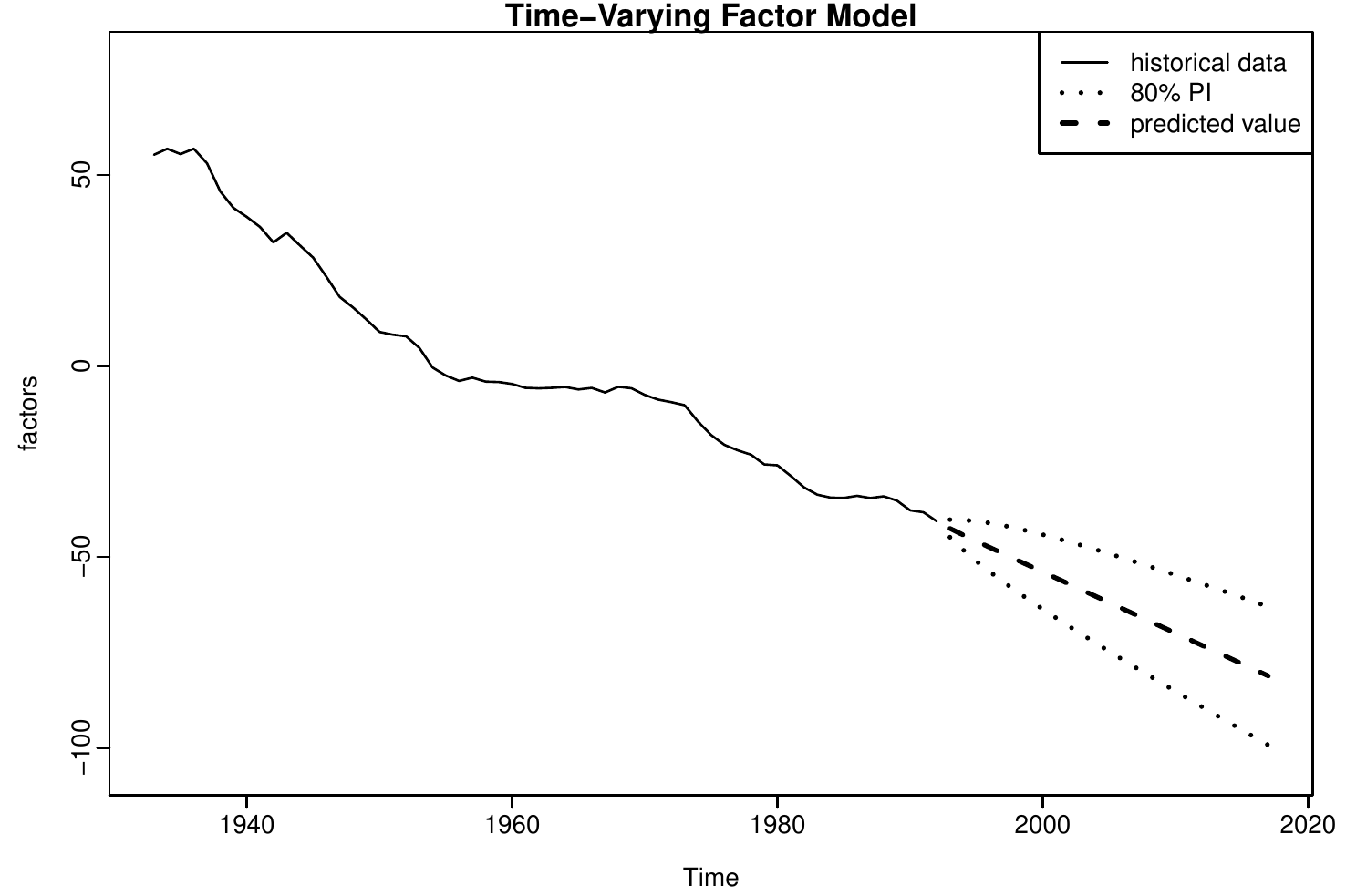}
	\caption{Out-of-sample forecast of the common factor, with the model fitted on $1933$ to $1999$ and the forecast horizon over $1993$ to $2017$; predicted value (dashed line), $80\%$ PI (dotted line)
	}
	\label{f5}
\end{figure}

Figure \ref{f5} plots the historical and predicted values of the common factor of the time varying factor model along with the associated $80\%$ prediction intervals, which is based on the ARIMA model fitted in Section \ref{subsec: fit}. The dashed downward line shows that the latent factor will keep declining in the future, and there is an $80\%$ chance that a future observation will be covered by the corresponding prediction interval (represented by area between the red dashed lines). 

 Since the time-varying factor model and the Lee-Carter model have similar common factor and the corresponding fitted ARIMA model, the forecasts of the common factor are close to each other too. Hence, the major difference of prediction accuracy between the time-varying factor model and the Lee-Carter model lies in the factor loadings.  We extrapolate the factor loadings obtained from the time-varying factor model using both the naive method and local linear regression introduced in Section \ref{subsec: forecast}, respectively. We then forecast the mortality rates into the future using both the time-varying and Lee-Carter models.

\begin{figure}[hbt!]
	\begin{minipage}{0.45\textwidth}
		\includegraphics[width=\linewidth, page=1]{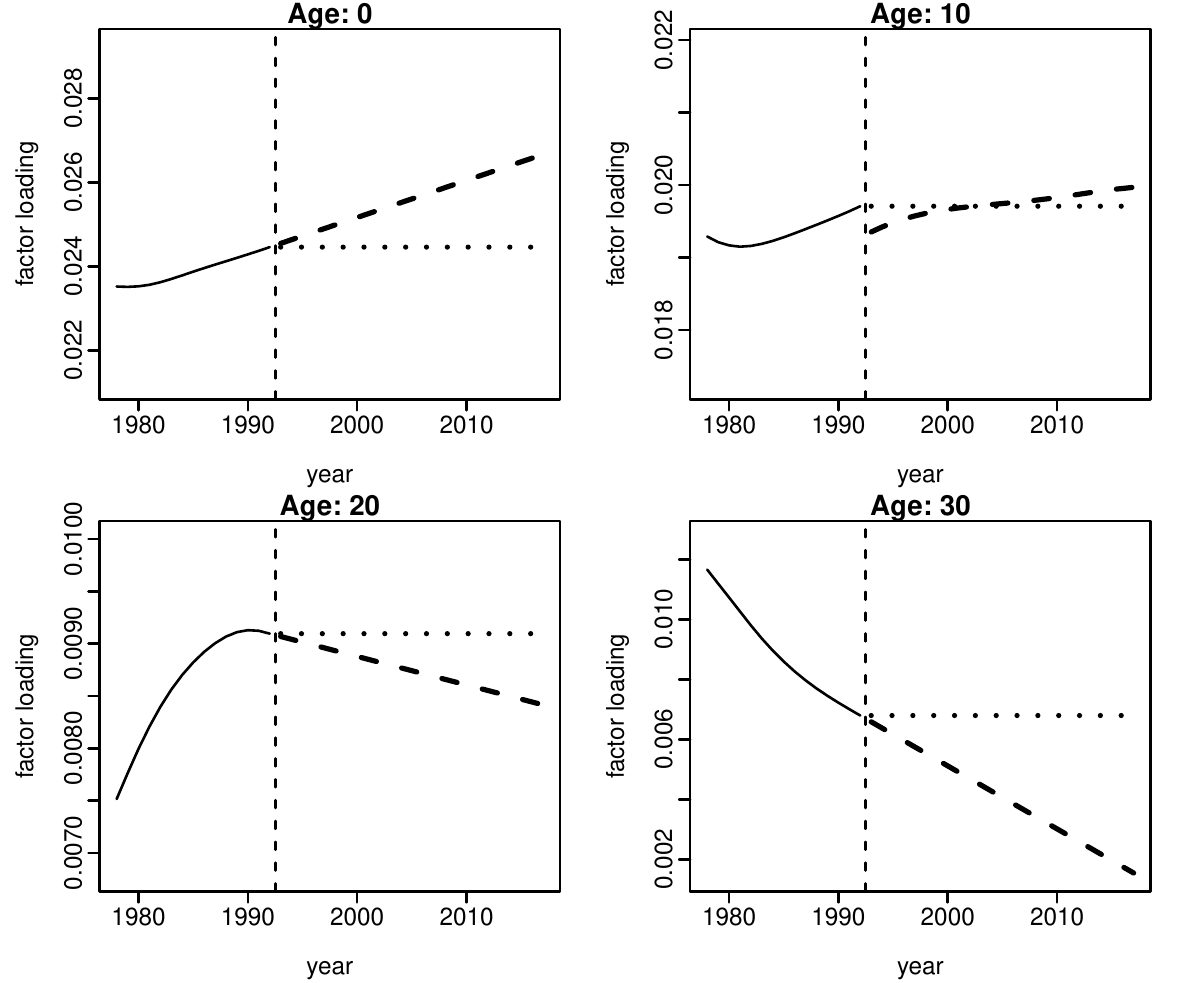}
		\includegraphics[width=\linewidth, page=2]{fig/f6_1.pdf}
		\includegraphics[width=\linewidth, page=1]{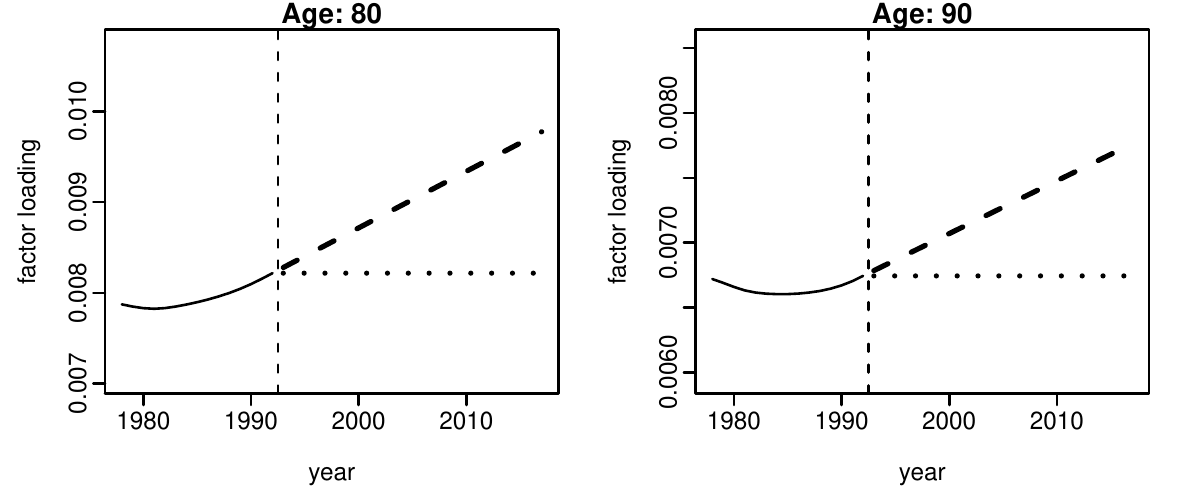}
		\caption{Plots of the estimated and extrapolated factor loadings based on naive method (dotted lines) \& local regression method (dashed lines) for age $0, 10, \dots, 90$.
		}
		\label{f6}
	\end{minipage}%
	\hfill
	\begin{minipage}{0.45\textwidth}
		\includegraphics[width=\linewidth, page=1]{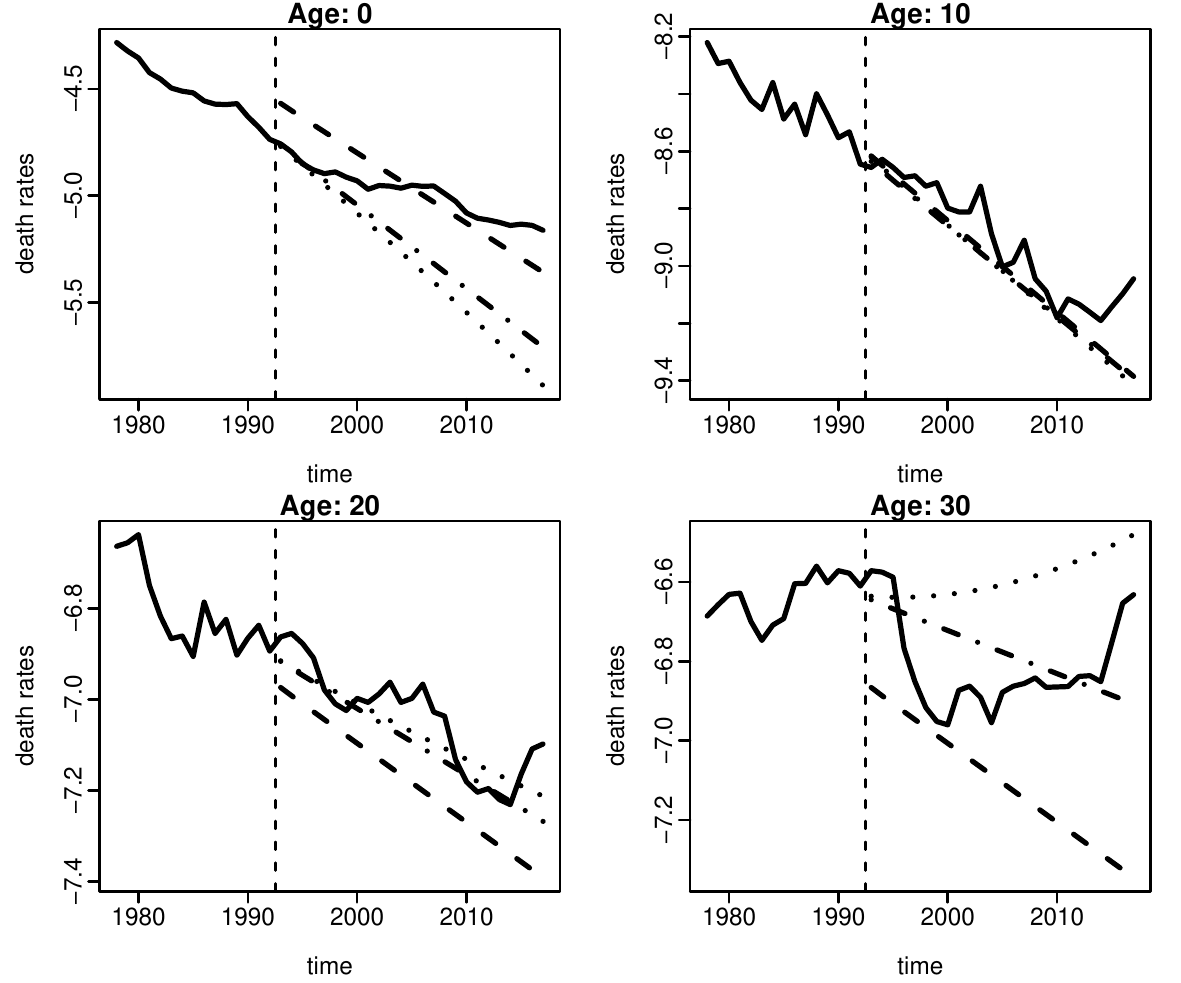}
		\includegraphics[width=\linewidth, page=2]{fig/f7_1.pdf}
		\includegraphics[width=\linewidth, page=1]{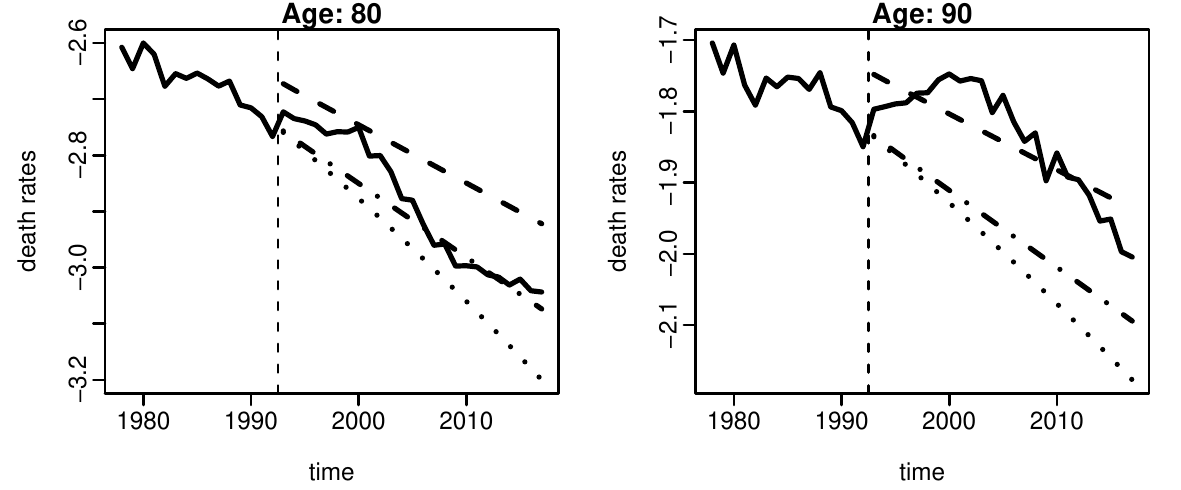}
		\caption{The actual data (solid line) versus the predicted values from the time-varying model (naive method: dash-dotted line; local linear regression: dotted line) and the Lee-Carter model (dashed line); the data have been log-transformed.
		}
		\label{f7}
	\end{minipage}
\end{figure}

Figure \ref{f6} plots the estimated and extrapolated factor loadings of the time varying factor models. Figure \ref{f7} plots the actual data and predicted values using the three above-mentioned methods. From Figure \ref{f6}, we see that the  local regression method follows the recent historical trend of factor loadings, while the naive method stays at a constant level. Theoretically, if future factor loadings do not deviate significantly from the recent historical trend, the local linear regression may perform better than the naive method in forecasting. However, it may only be reasonable to assume that factor loadings will follow the local trends in the short-term. For long-term forecasting, this assumption is less reliable and the non-parametric forecasting method would lead to inferior results. In Section \ref{subsec: comparison}, we observe similar results in other countries' mortality forecasting. Hence, the naive method, with time-invariant forecasted factor loadings, is more suitable for long-term forecasting. And it may also be suitable for short-term forecasting if the long-term trend is consistent with the short-term trend. Since the benefit of using the local regression method decreases as the forecasting horizon increases, it is worthwhile to ask whether an optimal forecasting horizon exists for using the local regression method. This question will be answered in Section \ref{subsec: change}.

 The empirical analysis suggests that, the time-varying factor model (based on the naive method) performs better than the Lee-Carter model over the entire forecasting horizon ($1993$-$2017$). Using the mean squared prediction error (MSPE) to evaluate the out-of-sample forecasting performance, we see that the overall MSPE for the Lee-Carter model is $0.03085$, while the overall MSPE for the time-varying factor model (based on the naive method) is only $0.01804$. However, if we choose to use the local linear regression to extrapolate factor loadings, the time-varying factor model performs worse than Lee-Carter model, with the MSPE being $0.04768$. 
 
\begin{figure}[hbt!]
	\centering
	\includegraphics[width=0.7\linewidth]{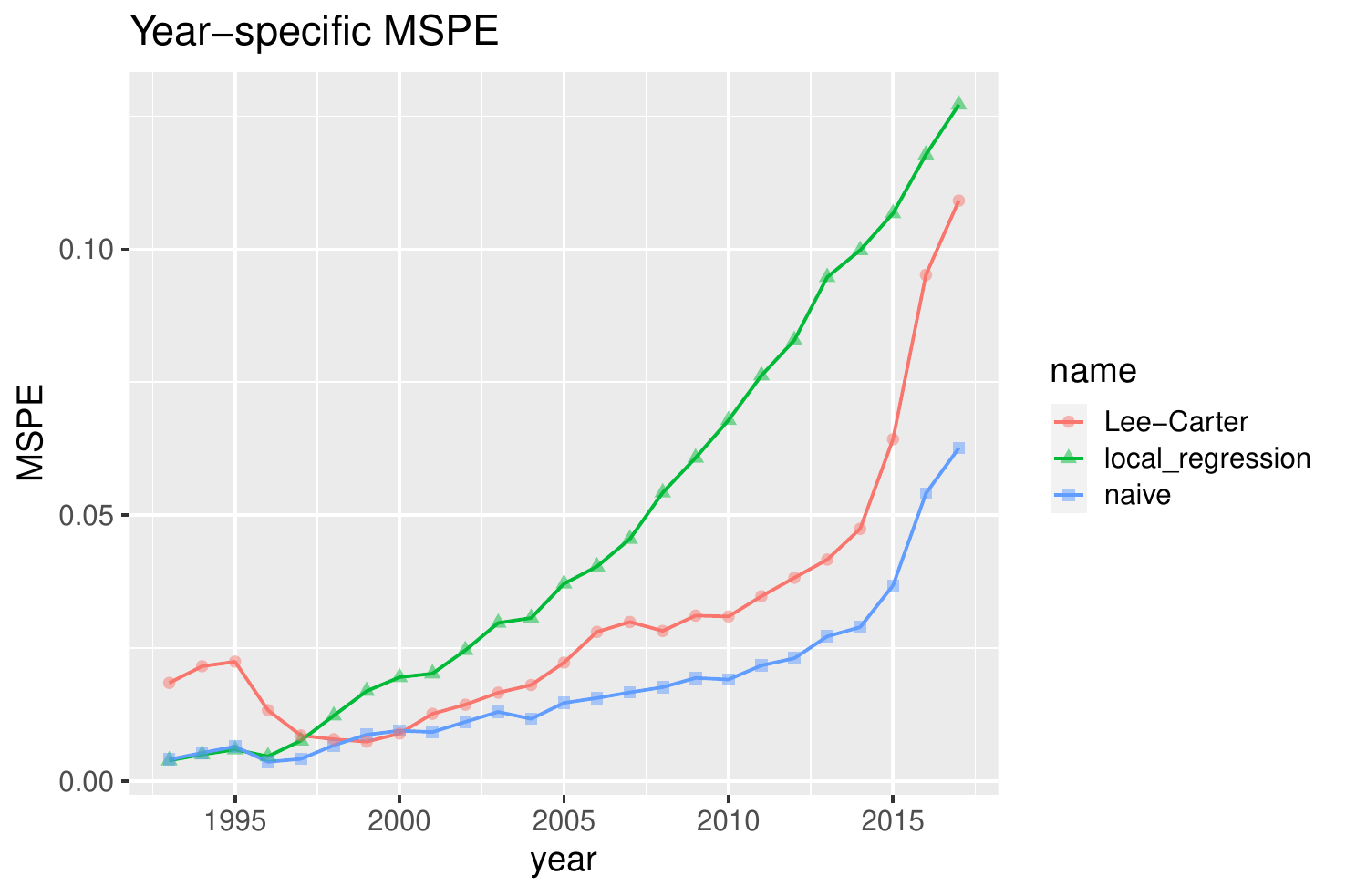}
	\caption{US: Year-specific MSPE for the time-varying model and the Lee-Carter model over $1993$ to $2017$; for time-varying model, both the naive method and the local regression method are used}
	\label{f8}
\end{figure}

 Figure \ref{f8} shows the year-specific MSPE for the time-varying factor models and the Lee-Carter model over the forecasting horizon $1993$ to $2017$. The year-specific MSPE is computed by averaging MSPE over all ages for each forecasting year. From Figure \ref{f8}, we can see that for the majority years, the MSPE of the time-varying model with the naive forecasting method is always the smallest one. From Section \ref{subsec: change}, we can see the reason is that the optimal 'boundary' between short-term and long-term forecasting in this case is estimated to be $0$. So the the time-varying model with the naive forecasting method is the best for both short-term and long-term forecasting. The MSPE for all the three methods are generally increasing over the years, as it is harder to forecast the farther future. We also notice that  the time-varying factor model based on the local regression method works better than the Lee-Carter model over 1993 to 1995. However, it has the worst performance for long-term forecasting. The time-varying model based on local regression method assumes the factor loadings change over time in the future, but it can only extend the recent trend, which may not be suitable for long-term forecasting. On the other hand, the Lee-Carter model and the time-varying model based on the naive method extrapolate factor loadings into the future as constants, which are usually more suitable for long-term forecasting.
 
\begin{figure}[hbt!] 
	\centering
	\includegraphics[width=0.7\linewidth]{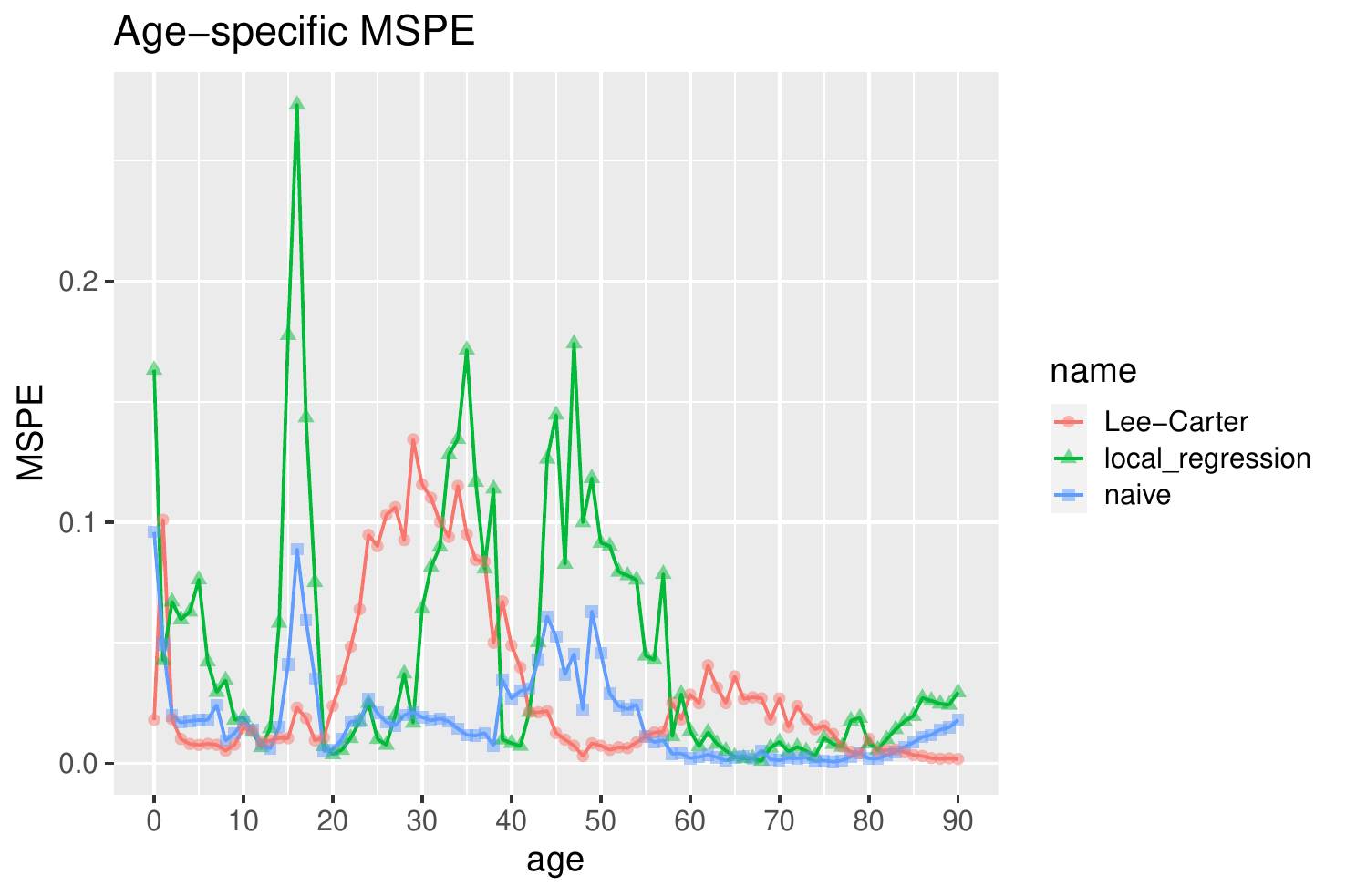}
	\caption{US: Age-specific MSPE for both the time-varying model and the Lee-Carter model; for time-varying model, both naive method and local regression method are used
	}
	\label{f9}
\end{figure}

Next, we investigate the forecasting performance of the time-varying factor model at different ages. Figure \ref{f9} shows the age-specific MSPE for the time-varying factor models and the Lee-Carter model. The square-dashed lines and triangle-dotted lines represent the MSPE of the time-varying factor models with the naive and local regression methods for each age, respectively, and the circle-solid line represents the MSPE of the Lee-Carter model.  We find that the naive forecasting method based on the time-varying factor model is almost always better than the local regression method for any age in this data. And roughly speaking, no matter which extrapolation method we choose to use, the time-varying models provide more accurate forecasts than the Lee-Carter model for age groups $20\sim40$ and  $60\sim80$. However, for the age group $40\sim60$, the forecasting performance of the time-varying factor models is worse than that of the Lee-Carter model. Thus, even though by using naive extrapolation method  the time-varying factor model improves the overall performance (over $40\%$ in terms of the MSPE) significantly, it cannot outperform the Lee-Carter model for some ages. The main advantage of the time-varying model is to forecast mortality rates for the young adulthood ($20\sim40$) and the older adulthood ($60\sim80$).

\subsection{Model Comparisons for Multiple Countries}
\label{subsec: comparison}

We apply and compare different models using mortality data of multiple countries. In particular, the functional data model proposed by \citet{hyndman2007robust} is also considered for comparison purposes and we call it `Hyndman-Ullah model' in the rest of the paper. It is a multi-factor extension of the Lee-Carter model, allowing for multiple age-time interaction terms to capture the complex structure of the data. Similar to the Lee-Carter model (i.e. the classical factor model with one factor), it is also commonly considered as a benchmark for mortality forecasting. To determine the order $K$ of Hyndman-Ullah model (similar to the number of factors $R$ in this paper), we choose the best value of $K$ by minimizing Integrated Squared Forest Error. Please refer to \citet{hyndman2007robust} for more details. 

In Table \ref{t2}, we present the results of the overall MSPE for different countries, forecast horizons and methods. We use the longest available dataset for training purposes, with different forecasting horizons listed in Table \ref{t2}. 
To investigate the performance of the short-term and long-term forecasting, we consider multiple forecasting horizons with different lengths, including $5$, $10$, $15$, $20$ and $25$ years. Note that, as the number of total historical years is fixed for each country, the number of training years changes with the length of the forecasting horizon. For example, the total historical years for Australia are 1921$\sim$2018 (98 years). Therefore, the corresponding training years for forecasting horizons 2014$\sim$2018 and 2009$\sim$2018 are 1921$\sim$2013 (93 years) and 1921$\sim$2008 (88 years), respectively. Please refer to Table \ref{t1} for the available historical years for each country.
The number of factors is always estimated as 1 ($\widehat{R} = 1$) for both the classical factor model the time-varying model. Hence, we use `Lee-Carter' to represent the classical factor model (with $R=1$) in this section. As for Hyndman-Ullah model, we record the estimated order $K$ for each scenario in Table \ref{t2}.

{\footnotesize
\begin{longtable}{c *{8}{Y}}
\caption{Overall MSPE by country, forecast horizon and method (\textbf{Hyndman-Ullah}: functional data model in \citet{hyndman2007robust} (estimated order K is shown in parenthesis for each scenario), \textbf{Lee-Carter}: classical factor model, \textbf{TV-local regression}: time-varying factor model based on local linear regression, \textbf{TV-naive}: time-varying factor model based on naive method)} \label{t2} \\

\hline \multicolumn{1}{c}{Country} & \multicolumn{1}{c}{Forecast Horizon} & \multicolumn{1}{c}{Hyndman-Ullah} & \multicolumn{1}{c}{Lee-Carter} & \multicolumn{1}{c}{TV-Local Regression} & \multicolumn{1}{c}{TV-Naive}\\ \hline 
\endfirsthead

\multicolumn{8}{c}%
{\tablename\ \thetable{} -- continued from previous page} \\
\hline \multicolumn{1}{c}{Country} & \multicolumn{1}{c}{Forecast Horizon} & \multicolumn{1}{c}{Hyndman-Ullah} & \multicolumn{1}{c}{Lee-Carter} & \multicolumn{1}{c}{TV-Local Regression} & \multicolumn{1}{c}{TV-Naive} \\ \hline 
\endhead

\hline \multicolumn{8}{c}{{Continued on next page}} \\ \hline
\endfoot

\hline
\endlastfoot

\textbf{Australia}   & $2014\sim2018$  & $\bf{0.012}$ (K = 8) & $0.037$ & $0.013$ & $0.014$ \\ 
   & $2009\sim2018$  & $0.022$ (K = 5) & $0.049$ & $\bf{0.016}$ & $0.019$ \\ 
   & $2004\sim2018$  & $0.036$ (K = 10) & $0.064$ & $\bf{0.026}$ & $0.028$ \\ 
   & $1999\sim2018$  & $0.082$ (K = 3) & $0.094$ & $0.045$ & $\bf{0.042}$ \\ 
   & $1994\sim2018$  & $0.053$ (K = 2) & $0.112$ & $0.089$ & $\bf{0.043}$ \\ 
\hline
\textbf{Canada}   & $2012\sim2016$  & $\bf{0.012}$ (K = 8) & $0.044$ & $0.015$ & $0.016$ \\
   & $2007\sim2016$ & $0.018$  (K = 3) & $0.045$ & $\bf{0.016}$ & $0.018$ \\ 
   & $2002\sim2016$ & $0.021$  (K = 9) & $0.049$ & $\bf{0.019}$ & $0.020$ \\ 
   & $1997\sim2016$ & $0.021$  (K = 6) & $0.051$ & $0.025$ & $\bf{0.020}$ \\ 
   & $1992\sim2016$ & $0.032$  (K = 6) & $0.061$ & $0.038$ & $\bf{0.024}$ \\ 
\hline
\textbf{France}  & $2013\sim2017$ & $\bf{0.009}$ (K = 5) & $0.047$ & $0.0146$ & $0.0154$ \\ 
& $2008\sim2017$ & $\bf{0.016}$  (K = 10) & $0.054$ & $0.021$ & $0.022$ \\ 
& $2003\sim2017$ & $\bf{0.042}$  (K = 8) & $0.081$ & $0.049$ & $0.047$ \\ 
& $1998\sim2017$ & $\bf{0.059}$  (K = 3) & $0.101$ & $0.068$ & $0.061$ \\ 
& $1993\sim2017$ & $\bf{0.079}$  (K = 8) & $0.128$ & $0.096$ & $0.085$ \\ 
\hline
\textbf{Italy}  & $2013\sim2017$ & $\bf{0.012}$ (K = 9) & $0.044$ & $0.014$ & $0.015$ \\
& $2008\sim2017$ & $\bf{0.018}$  (K = 4) & $0.057$ & $0.028$ & $0.029$ \\ 
& $2003\sim2017$ & $\bf{0.047}$  (K = 10) & $0.084$ & $0.064$ & $0.052$ \\ 
& $1998\sim2017$ & $0.090$  (K = 5) & $0.098$ & $0.077$ & $\bf{0.058}$ \\ 
& $1993\sim2017$ & $0.099$  (K = 6) & $0.108$ & $0.073$ & $\bf{0.067}$ \\ 
\hline
\textbf{Japan}   & $2014\sim2018$  & $0.012$ (K = 3) & $0.031$ & $\bf{0.009}$ & $\bf{0.009}$ \\ 
& $2009\sim2018$  & $\bf{0.010}$  (K = 6) & $0.034$ & $0.017$ & $0.011$ \\ 
& $2004\sim2018$  &  $0.039$  (K = 4) & $0.072$ & $0.023$ & $\bf{0.016}$ \\ 
& $1999\sim2018$  & $0.024$  (K = 3) & $0.055$ & $0.018$ & $\bf{0.017}$ \\ 
& $1994\sim2018$  & $\bf{0.030}$  (K = 9) & $0.072$ & $0.037$ & $0.034$ \\ 
\hline
\textbf{USA}  & $2013\sim2017$   & $\bf{0.011}$ (K = 8) & $0.028$ & $0.013$ & $0.014$ \\ 
  & $2008\sim2017$   & $\bf{0.010}$  (K = 10) & $0.024$ & $0.015$ & $0.014$ \\ 
  & $2003\sim2017$  & $0.017$  (K = 6) & $0.025$ & $0.024$ & $\bf{0.016}$ \\ 
  & $1998\sim2017$  & $0.111$  (K = 10) & $0.029$ & $0.031$ & $\bf{0.021}$ \\ 
  & $1993\sim2017$ & $0.026$  (K = 10) & $0.031$ & $0.048$ & $\bf{0.018}$ \\ 
\end{longtable}
}

From Table \ref{t2}, we see that the time-varying models can significantly improve the out-of-sample forecasting performance compared with the Lee-Carter model. The time-varying factor models and Hyndman-Ullah model perform the best in most cases. The Hyndman-Ullah model is especially suitable for mortality forecasting of the France data, as shown in \citet{hyndman2007robust}. In addition,  for the time-varying factor models, the local regression method tends to perform better for short-term forecasting,  while the naive method is better for long-term forecasting. Compared with all the other models, the time-varying model based on the naive method show superior results for long-term forecasting, while the Lee-Carter model always has the worst forecasting performance. 

\begin{figure}[hbt!]
	\centering 
	\begin{minipage}[c]{0.45\textwidth}
		\includegraphics[width=\linewidth, page=1]{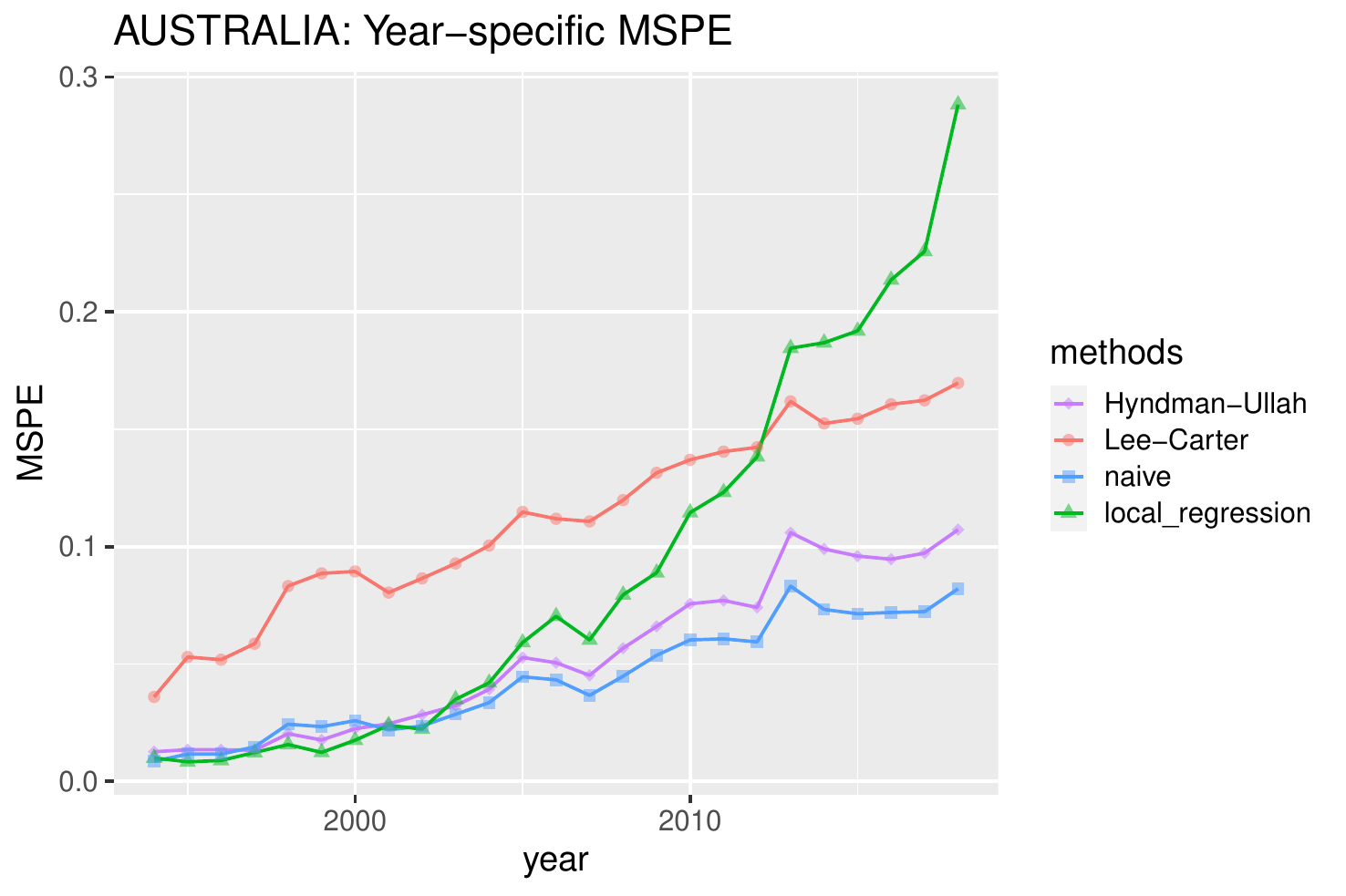}
		\includegraphics[width=\linewidth, page=3]{fig/f10.pdf}
		\includegraphics[width=\linewidth, page=5]{fig/f10.pdf}
	\end{minipage}
	\hspace*{0.05cm}
	\begin{minipage}[c]{0.45\textwidth}
		\includegraphics[width=\linewidth, page=2]{fig/f10.pdf}
		\includegraphics[width=\linewidth, page=4]{fig/f10.pdf}
		\includegraphics[width=\linewidth, page=6]{fig/f10.pdf}
	\end{minipage}
	\caption{Year-specific MSPE by country and method (\textbf{Hyndman-Ullah}: functional data model in \citet{hyndman2007robust}, \textbf{Lee-Carter}: classical factor model, \textbf{local regression}: time-varying factor model based on local linear regression, \textbf{naive}: time-varying factor model based on naive method); length of forecast horizon is 25 years}
	\label{f10}
\end{figure}

In Figure \ref{f10}, by fixing the length of the forecast horizon to be 25 years, we plot the year-specific MSPE for all the countries using different forecasting methods. Roughly speaking, the time-varying model based on the naive method always has the most accurate forecasts, while the Lee-Carter model always performs the worst. Additionally, the time-varying model based on the local regression method usually performs well (similar to or better than the naive method) at the first few years, while it deteriorates as time goes by. For mortality data of Australia and the US, it has worse forecasting performances than the Lee-Carter model in the long term.

\subsection{Estimate the Optimal `Boundary'}
\label{subsec: change}
From the empirical results above, we see that under the framework of time-varying model, the local regression method (by assuming factor loading will change in the future) is better at short-term forecasting while the naive method (by assuming factor loading is in-variant in the future) is better at long-term forecasting. Therefore, we are interested in the boundary between short-term and long-term forecasting that divides the forecast horizon according to the predictive power of the local regression method and the naive method. 

By applying the estimation method introduced in Section \ref{sec: choice}, we estimate the optimal `boundary' between the short-term and  long-term forecasting, which is favoured by the local regression method (time-varying forecast of the factor loading) and the naive method (time-invariant forecast of the factor loading) respectively. Recall that the optimal `boundary' can be regarded as the optimal value of $k$ defined in the hybrid forecasting method in Section \ref{sec: choice}.
Applying the same datasets introduced in Section \ref{sec: data}, we use the last 25 years of the historical data as the validation set, and the remaining data as the training set. In addition, to check the sensitivity of the least squares estimator to the division of the validation and testing sets, we consider the last $p$ years ($p=15,20,25,30$, respectively) of data as the validation set and the remaining data as the training set. The results are shown in Appendix B. 

\begin{figure}[hbt!]
	\centering
	\begin{minipage}[c]{0.35\textwidth}
		\includegraphics[width=\linewidth, page=1]{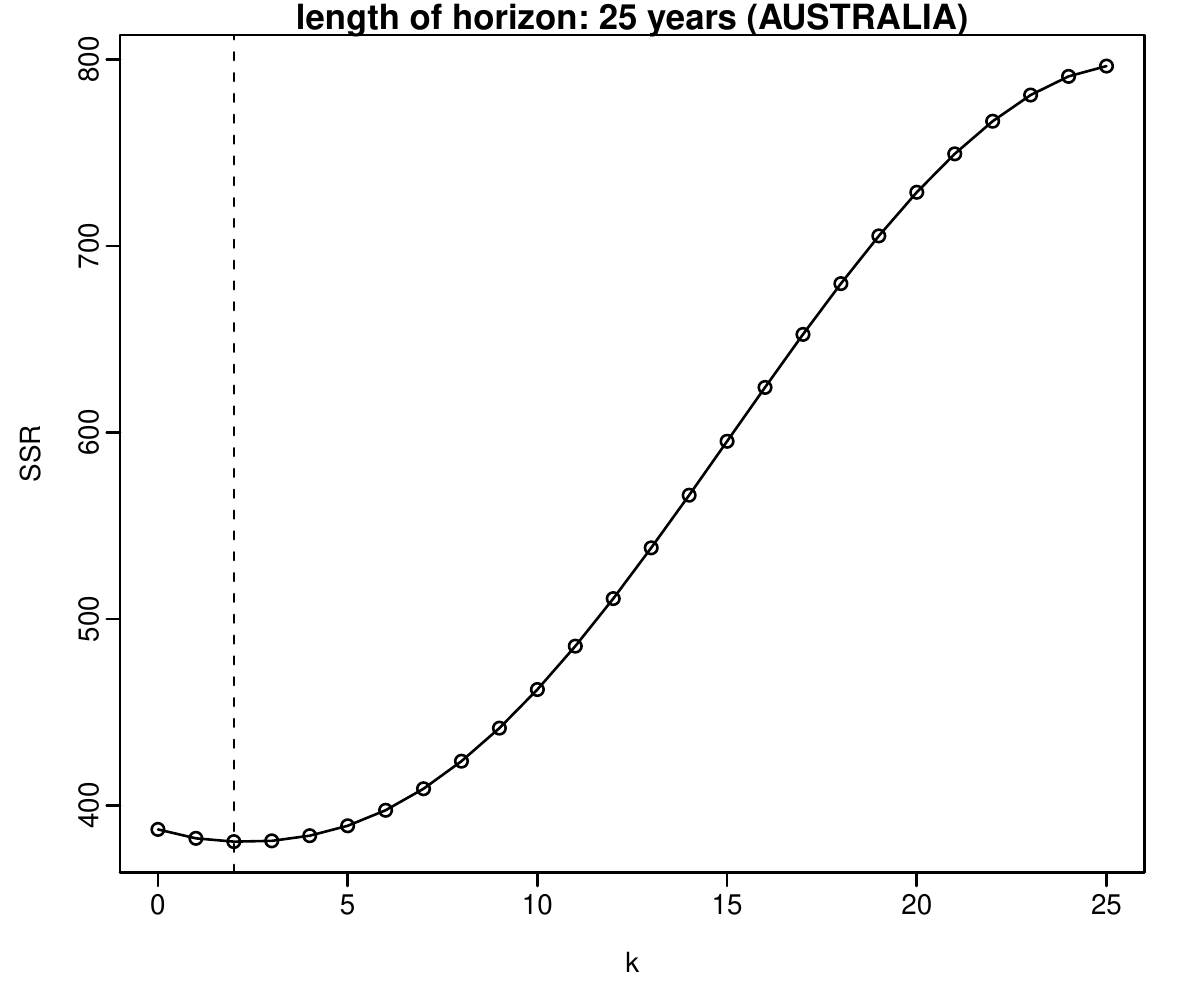}
		\includegraphics[width=\linewidth, page=3]{fig/f11.pdf}
		\includegraphics[width=\linewidth, page=5]{fig/f11.pdf}
	\end{minipage}
	\hspace*{0.05cm}
	\begin{minipage}[c]{0.35\textwidth}
		\includegraphics[width=\linewidth, page=2]{fig/f11.pdf}
		\includegraphics[width=\linewidth, page=4]{fig/f11.pdf}
		\includegraphics[width=\linewidth, page=6]{fig/f11.pdf}
	\end{minipage}
	\caption{Plots of the total sum of squared residuals (SSR) versus the length ($k$) of the short-term forecast horizon (based on the hybrid forecasting method of time-varying factor model); length of forecast horizon: $25$}
	\label{f11}
\end{figure}

As defined in Section \ref{sec: choice}, for each set of data we compute the least squares estimator for the optimal `boundary' in the forecasting horizon as 
\begin{align*}
\widehat{k}= \underset{0\le k \le T-T_0}{argmin}SSR(k),
\end{align*}
where $k=0, 1, 2, \dots, T-T_0$. Here $k$ represents the length of the short-term forecasting horizon, or the `boundary' between short-term (based on the local regression method) and long-term (based on the naive method) forecasting. The local linear regression is used to make forecasts from $T_0+1$ to $T_0+k$; while the naive method (i.e. keep the factor loading the same as that in time $T_0+k$) is used to make forecasts from $T_0+k+1$ to $T$.

We plot the total SSR versus  $k$ in Figure \ref{f11}.  As shown in the plots, for each country, there is a minimum point of $k$ corresponding to the smallest values of SSR, which indicates the optimal length of the short-term forecasting horizon for the time-varying factor model. For example, the plot of Italy shows that, when the value of $k$ equals 7, SSR achieves the smallest value. Thus, based on the historical Italy mortality data, we suggest that it is better to use the local regression method for the short-term forecasting (less than 7 leads), as it puts more weights on the recent observations. As for forecasting more than $7$ years ahead, the naive method  generates more accurate predictions. However, as for the US data, the optimal length of the short-term forecasting horizon is 0, which means there is no need to extrapolate factor loadings using local linear regression and we should use the naive method alone. In Figure \ref{f11}, we can observe jumps at $k=1$ for some countries (such as Canada and France).  When $k=0$, the hybrid method is just the naive method and the extrapolation of factor loading is based on the historical estimation; while when $k>1$, the hybrid method is based on the local regression method before the time $T_0+k$. 
Therefore, a sudden change of SSR when $k$ increases from $0$ to $1$ indicates that the pattern of factor loadings has a relatively large change after the time $T_0$. 

Now we have observed the existence of positive optimal `boundary' for some countries (such as Australia, Italy and Japan). Then based on the estimation of optimal `boundary' $k$, we can compare the out-of-sample forecasting performance of the hybrid method with other methods. Here, we consider the Australian mortality data as an example since the estimation of the optimal boundary for Australian data is relatively stable (see details in Remark \ref{rmk: 4} and Appendix B).

\begin{figure}[hbt!]
	\centering
	\includegraphics[width=0.7\linewidth]{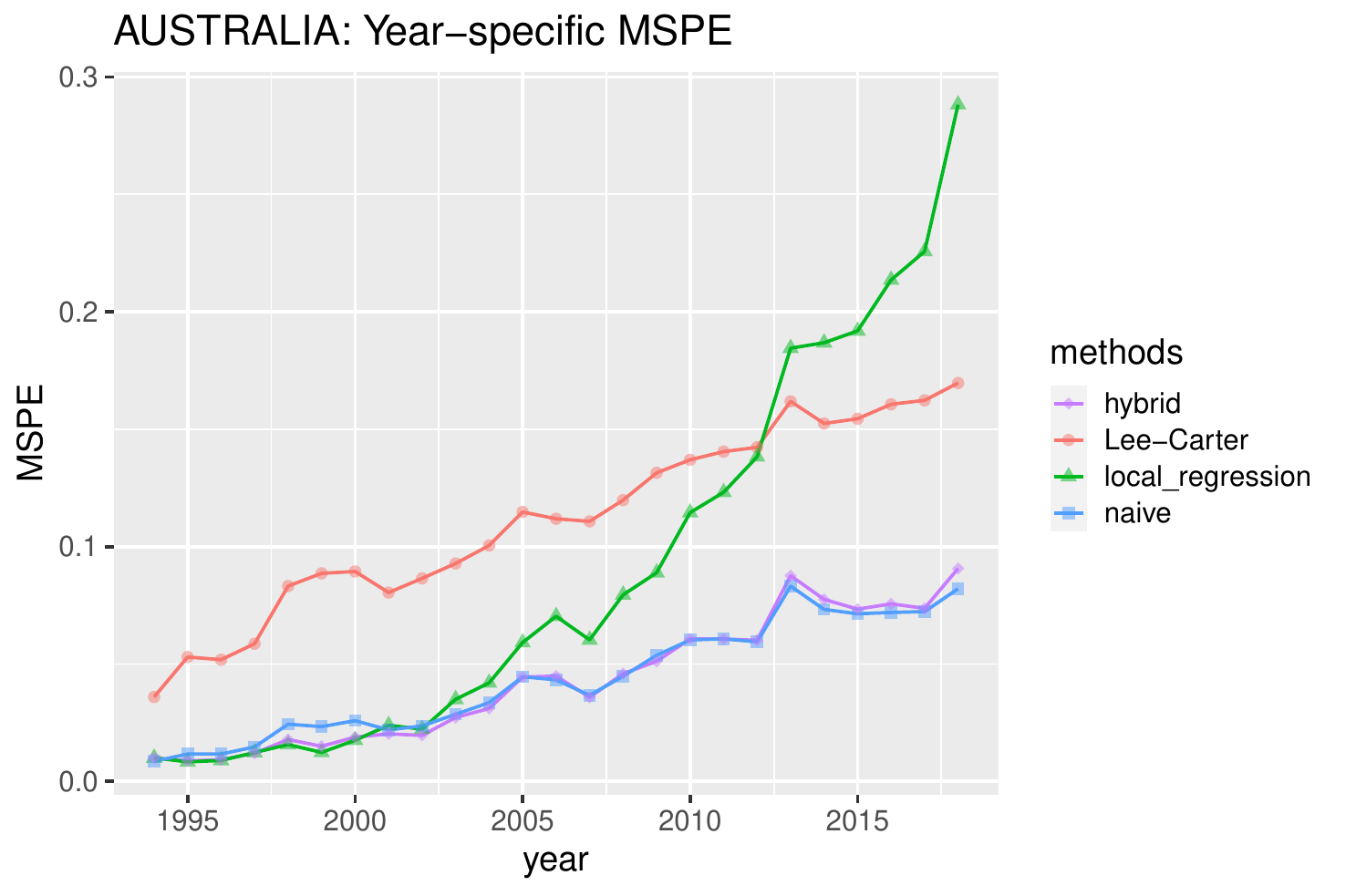}
	\caption{Australia: Year-specific MSPE over $1994$ to $2018$ (\textbf{Lee-Carter}: classical factor model, \textbf{local regression}: time-varying factor model based on local regression method), \textbf{naive}: time-varying factor model based on naive method), \textbf{hybrid}: time-varying factor model based on hybrid method)}
	\label{f12}
\end{figure}

Similar to previous analysis, we choose the last 25 years of the historical data as the testing set. And the remaining data is the training set. To estimate $k$, we use the last 25 years of the training data as the validation set to estimate the optimal `boundary'. We then apply the estimation procedure in Section \ref{sec: choice} to the training data and the forecasting method in the testing data  using the estimated optimal `boundary' to get the out-of-sample performance. The empirical results are shown in Figure \ref{f12} and Table \ref{t3}. 

In Table \ref{t3}, we compute the overall MSPE over $1994$ to $2018$ for each forecasting method. We find that the time-varying model based on the hybrid forecasting method has the best out-of-sample performance among all four methods. And the naive method is a little bit worse than the hybrid method. Additionally, in  Figure \ref{f12}, we plot the year-specific MSPE for all different methods. For short-term forecasting, the hybrid method shows similar performance with the local regression method; while for the long-term forecasting, the hybrid method shows similar performance with the naive method.  Therefore, the hybrid method is superior for both short-term and long-term forecasting, as it  benefits from the advantages of both methods. In practice, both naive method and hybrid method are recommended. The hybrid method could produce more accurate predictions while the naive method is much easier to implement. 

\begin{table}[!htbp] \centering 
	\small
	\caption{Australia: Overall MSPE over $1994$ to $2018$ (\textbf{Lee-Carter}: classical factor model, \textbf{TV-Local Regression}: time-varying factor model based on local regression method), \textbf{TV-Naive}: time-varying factor model based on naive method), \textbf{TV-Hybrid}: time-varying factor model based on hybrid method)} 
	\label{t3} 
	\begin{tabularx}{\textwidth}{c *{4}{Y}} 
		\toprule
		Country & Lee-Carter & TV-Local Regression & TV-Naive & TV-Hybrid \\
		\midrule 
		Australia & 0.11162 & 0.08932  & 0.04342  & 0.04288  \\
		\bottomrule 
	\end{tabularx} 
\end{table} 

\begin{rmk}
	\label{rmk: 4}
In Appendix B, we further investigate the sensitivity of the 
optimal `boundary' estimation  by showing  plots of the total sum of squared residuals (SSR) versus $k$ using different lengths of training and testing datasets. For some countries, like Japan and USA, the least squares estimators are relatively stable. However, for countries like Canada and Italy, the estimation of the optimal boundary is
somewhat sensitive to the length of the dataset.
This is reasonable as the time-varying factor model highly depends on the data and its intrinsic patterns.  Therefore, when using different training data and forecasting horizons to estimate optimal `boundary', we may obtain different values of $k$.

\end{rmk}

\section{Monte Carlo Simulations}
\label{sec: simulation}
In this section, we further investigate the prediction performance of the time-varying factor model and the classical factor model through Monte Carlo simulations. We use examples with different structures of the factor loadings to illustrate that the time-varying factor model can improve the forecasting accuracy when the `true' factor loadings change over time.
In addition, we explain under which conditions the naive method performs better than the local regression method even in the short-term.

Similar to the previous empirical analysis, we denote the classical factor model  as `Lee-Carter' since there is only one factor. We denote the two forecasting methods based on the time-varying factor model as `TV-Local Regression' method and `TV-Naive' method, respectively.
We show that when the `true' factor loadings change over time, both forecasting methods based on the time-varying factor model outperform the Lee-Carter method in forecasting. Moreover, the `TV-Naive' method performs similarly to the `TV-Local Regression' method for short-term forecasting; while it performs better than the `TV-Local Regression' for long-term forecasting.

\subsection{Data Generating Processes (DGP's)}
We generate the centered data $x_{i,t}$ with one common factor:
\begin{align*}
x_{i,t}=b_{i,t}\cdot{k_t}+\epsilon_{i,t}, \ \ i=1, 2, \ldots, N,
\end{align*}
where $k_t=k_{t-1}+w_t$. $w_t$ follows independent identically distributed normal distributions, $N(0,0.8^2)$. Thus the common factor $k_t$ follows a random walk. We consider the following settings for different factor loadings $b_{i,t}$ and error terms $\epsilon_{i,t}$. In each setting below, we apply the normalization condition mentioned in Section \ref{sec: model}, so that $b_{i,t}$ is normalized to sum to unity for each $t$.
\begin{itemize}
	\item \textbf{DGP 1} (time-invariant factor loading):

\begin{align*}
b_{i,t}= b_{i}\sim{i.i.d} \ {uniform(0,1)},\  \epsilon_{i,t}\sim{i.i.d} \ N(0,0.1^2).
\end{align*}

	\item \textbf{DGP 2} (single-point structural change):
	
	 For $i=1, 2, \cdots, N/2$,
\begin{align*} 
b_{i,t}=\left\{
        \begin{array}{ll}
           b_{i}  & \quad for\ t=1,2,...,T/2 \\
            b_i+1 & \quad for\ t=T/2+1,...,T
        \end{array}
    \right. ;
\end{align*}
and for $i=N/2+1, \cdots, N$,
\begin{align*}
b_{i,t}=\left\{
        \begin{array}{ll}
           b_{i}  & \quad for\ t=1,2,...,T/2 \\
            b_i-1 & \quad for\ t=T/2+1,...,T
        \end{array}
    \right. ;
\end{align*}
\begin{align*}
b_{i}\sim{i.i.d} \ {uniform(1.1,1.9)},\  \epsilon_{i,t}\sim{i.i.d} \ N(0,0.03^2).
\end{align*}

	\item \textbf{DGP 3} (continuous structural change):
\begin{align*}
b_{i,t}=\frac{1}{1+e^{(\frac{6i}{N}+2-\frac{12t}{T})}},\  \epsilon_{i,t}\sim{i.i.d} \ N(0,0.1^2).
\end{align*}
\end{itemize}

\textbf{DGP 1} follows the Lee-Carter model with time-invariant factor loadings, and \textbf{DGP 2} and \textbf{DGP 3} exhibit different structures of the time-varying factor loadings. \textbf{DGP 2} describes a single-point structural change in the factor loadings; while \textbf{DGP 3} considers a continuous structural change in the factor loadings. For each $i$, the factor loadings generated in \textbf{DGP 3} are monotonic functions and would converge to some constant as  time goes by. The factor loadings in \textbf{DGP 3} may go up and down and would never diverge to extreme values, which is similar to the estimated time-varying factor loadings in Figure \ref{f3} using the US mortality data.

\subsection{Comparison of the Forecasting Performance}
To compare the forecasting performance of the time-varying factor model and the Lee-Carter model (classical time-invariant factor model), we use the out-of-sample testing approach in the following analysis. For each DGP, we simulate 100 data sets with the dimension and sample sizes $N=T=100$. For each data set, we consider the first $k$ years of the data as the training set, and the remaining $T-k$ years of the data as the testing set ($k=70,75,80,85,90,95$, respectively). The model is first fitted using the training set, and then forecasted in the testing set. We employ the mean squared prediction error (MSPE) as the measure to evaluate performance of different models.

\begin{figure}[hbt!]
\centering
		\includegraphics[width=0.3\linewidth]{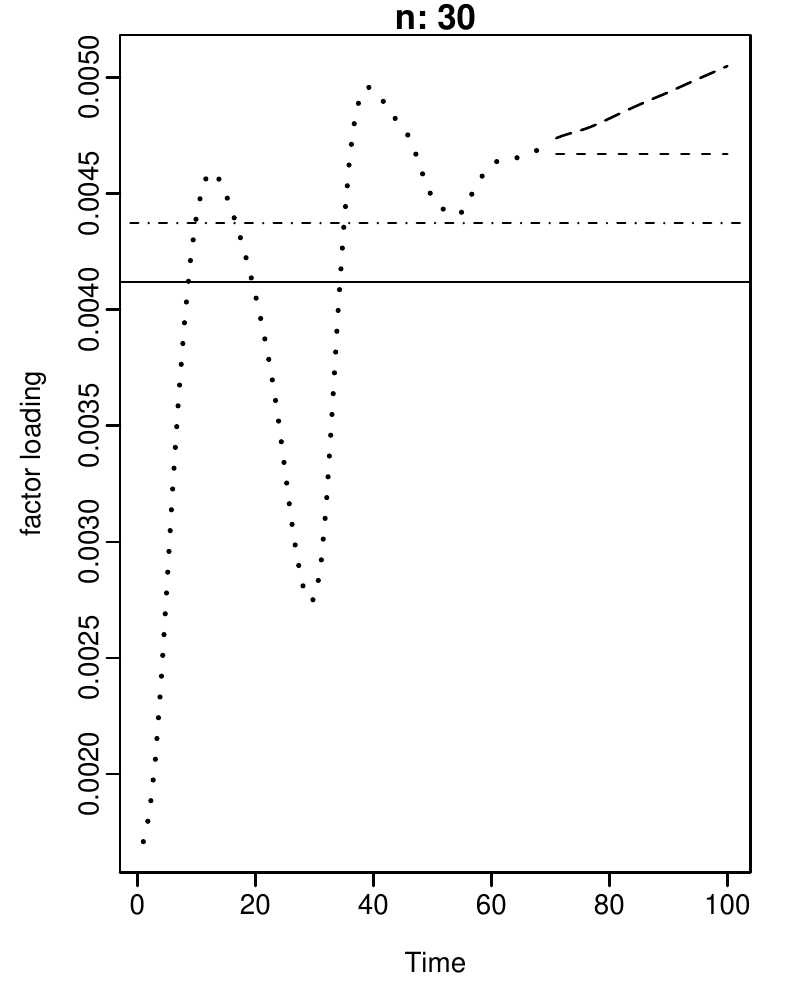}
		\includegraphics[width=0.3\linewidth]{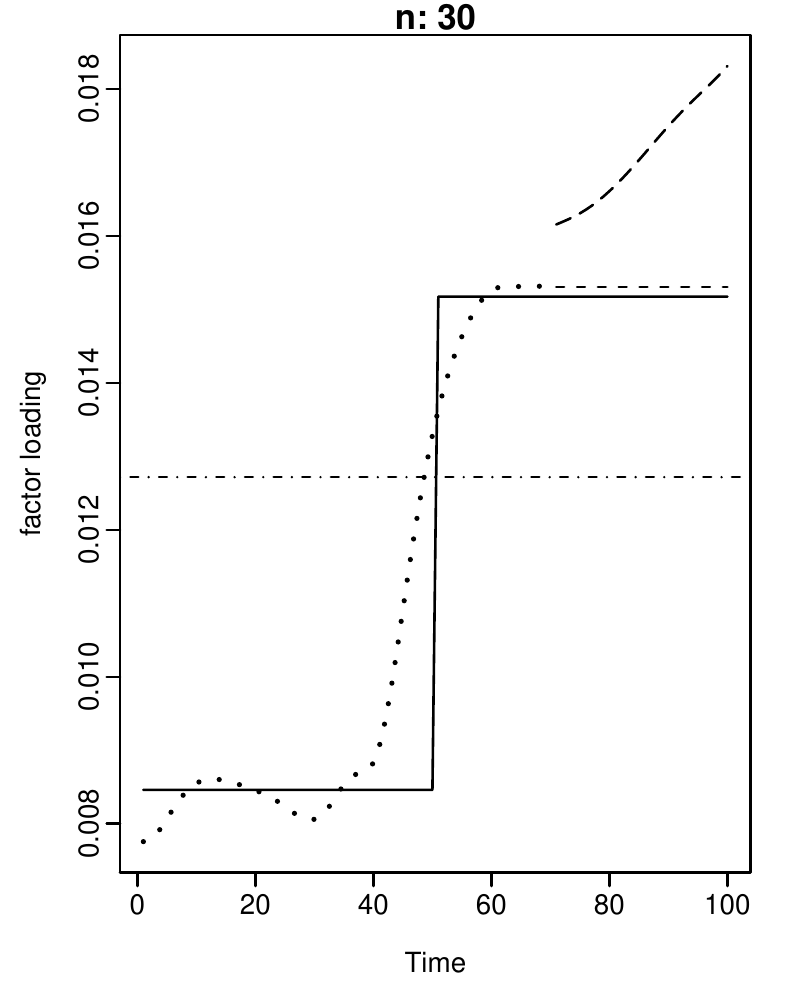}
		\includegraphics[width=0.3\linewidth]{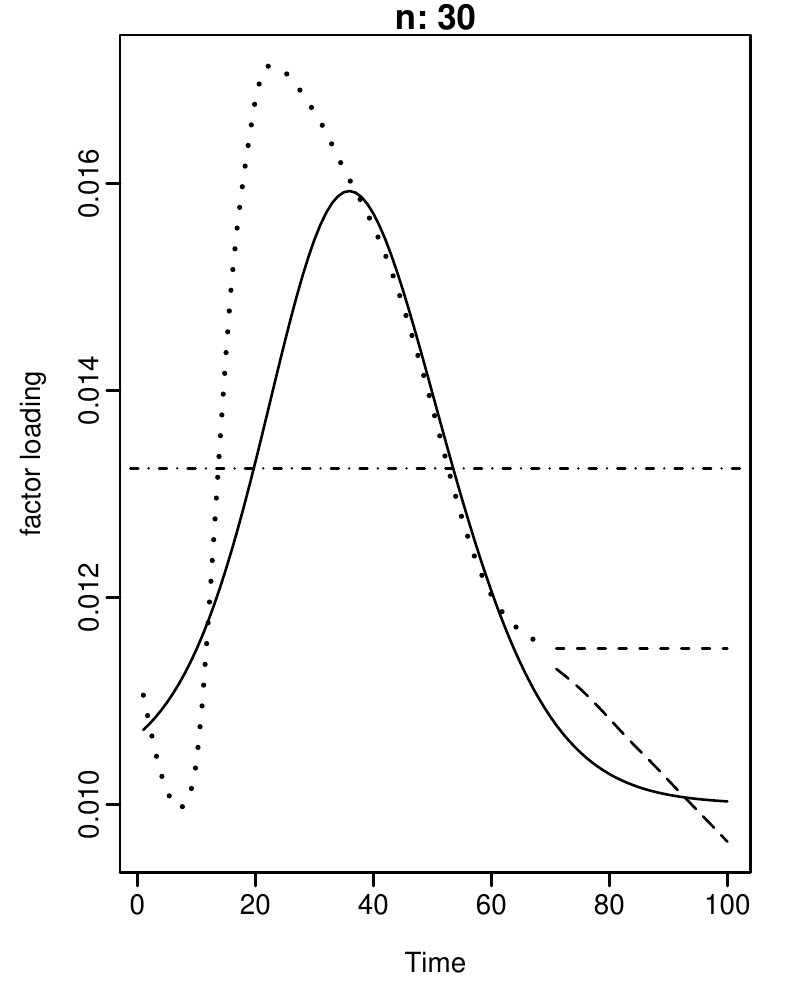}
		\caption{Comparison of the factor loadings: estimation and forecast. From left to right: \textbf{DGP 1}, \textbf{DGP 2}, \textbf{DGP 3}. $k = 70$. Solid line: true factor loadings. Dot-dashed line: estimation from the classical factor model (`Lee-Carter'). Dotted line: estimation from the time-varying factor model. Long dashed line: `TV-Local Regression'. Short dashed line: `TV-Naive'.}
		\label{simu1}
\end{figure}

Table \ref{s1} reports the comparison results based on various lengths of the training and testing sets.  An example of the estimated and forecasted factor loadings is shown in Figure \ref{simu1} to better explain the results.
As shown in Table \ref{s1}, the two time-varying methods perform worse than the Lee-Carter model for \textbf{DGP 1}, which assumes the time-invariant factor loadings. The less accurate prediction results can be attributed to the inaccurate estimation from the time-varying model, which is supported by the left plot of Figure \ref{simu1}. When the `true' factor loadings are time-invariant, the estimation from the time-varying model goes up and down randomly due to the over-fitting problem of the non-parametric estimating method. Therefore, the forecasting based on these estimation is not satisfied. However, the Lee-Carter method  provides a close estimation and better forecasting in this case.


\begin{table}[!htbp] \centering 
  \small
  \caption{Comparison of forecasting performance of the time-varying factor model and the Lee-Carter model (based on the different lengths of training sets)} 
  \label{s1} 
\begin{tabularx}{\textwidth}{cc *{6}{Y}} 
\toprule
DGPs & methods & 70 & 75 & 80 & 85 & 90 & 95 \\
\midrule 
 & TV-Local Regression& $0.2300$ & $0.1873$ & $0.1249$ & $0.0852$ & $0.0590$ & $0.0344$ \\ 
DGP 1 &TV-Naive & $0.2239$ & $0.1849$ & $0.1228$ & $0.0837$ & $0.0582$ & $0.0342$ \\ 
 & Lee-Carter & $\bf{0.2209}$ & $\bf{0.1839}$ & $\bf{0.1222}$ & $\bf{0.0829}$ & $\bf{0.0575}$ & $\bf{0.0336}$ \\ 
\midrule 
&TV-Local Regression& $0.2597$ & $0.2046$ & $0.1230$ & $0.0818$ & $0.0528$ & $0.0265$ \\ 
DGP 2&TV-Naive & $\bf{0.2291}$ & $\bf{0.1874}$ & $\bf{0.1227}$ & $\bf{0.0817}$ & $\bf{0.0527}$ & $\bf{0.0265}$ \\ 
&Lee-Carter & $0.2542$ & $0.2121$ & $0.1482$ & $0.0946$ & $0.0643$ & $0.0372$ \\ 
\midrule
&TV-Local Regression & $0.1757$ & $0.1384$ & $0.0987$ & $0.0737$ & $0.0506$ & $0.0365$ \\ 
DGP 3&TV-Naive & $\bf{0.1704}$ & $\bf{0.1319}$ & $\bf{0.0940}$ & $\bf{0.0724}$ & $\bf{0.0499}$ & $\bf{0.0362}$ \\ 
&Lee-Carter & $0.2078$ & $0.1759$ & $0.1456$ & $0.1162$ & $0.0917$ & $0.0748$ \\
\bottomrule 
\end{tabularx} 
\end{table} 

\textbf{DGP 2} and \textbf{DGP 3} follow the structures of time-varying factor models with abrupt and continuous changes in the factor loadings, respectively. From Table \ref{s1}, we see that both the two time-varying methods perform better than the Lee-Carter method when the `true' factor loadings are time-varying. In particular, the naive method performs the best in these two cases, especially for long-term forecasting. The superior forecasts of the two time-varying methods result from the more accurate estimation of the time-varying factor loadings, which can be seen from the middle and right plots in Figure \ref{simu1}. We find that the Lee-Carter model cannot capture the changed factor loadings as it assumes the factor loadings are time-invariant, while the time-varying model can estimate those changing factor loadings accurately and provides a solid foundation for the forecasting step. 

Further analysis of Table \ref{s1} suggests that the local regression method and the naive method preform similarly for the relatively short-term forecasting ($k = 85, 90, 95$), while for long-term forecasting ($k = 30, 25$), the naive method performs  better. This phenomenon can be explained by the plot of \textbf{DGP 3} in Figure \ref{simu1}. From the plot, we see that the forecast of the local regression method follows the trend of the estimated factor loadings. Therefore, when the trend remains the same in a short period of time, the forecast  of the local regression method is satisfying. However, as the trend of the `true' factor loadings changes, the forecast of the local regression method diverges away from the true values in the long term. This is the drawback of the non-parametric forecasting method. On the other hand, the constant factor loadings used in the forecasting procedure of the naive method guarantee that the factor loadings will not diverge dramatically and result in a better performance for long-term forecasting. In addition, although the `true' factor loading in the training set changes, if it is time-invariant in the forecasting horizon, the naive method is also better than the local regression method even in the short term (which is supported by the plot of \textbf{DGP 2} in Figure \ref{simu1}). The aforementioned analysis could be used to explain why the estimated optimal `boundary' of the US data is 0 in Section \ref{subsec: change}. From Figure \ref{f3} we see that in the first years of the forecasting horizon, the trends of the factor loadings are either different from that in the training set or remain flat. However, the local regression method cannot capture the unknown changing trends, so the naive method can outperform it even in the short term. In this case, the naive method not only captures the time-varying factor loadings in the estimation step, but also uses the constant factor loadings in the forecasting step.


\section{Conclusion}
\label{sec: conclusion}

This paper develops a time-varying factor model for mortality modelling. Two forecasting methods, the local linear regression and naive method, are used to extrapolate the factor loadings into the future. To understand the optimal forecasting horizon of the two forecasting methods, we propose an approach to estimate the optimal `boundary' between short-term and long-term forecasting, which is favoured by the local linear regression and the naive method, respectively. Empirical analysis on mortality forecasting of multiple countries demonstrates the advantages of the new model.   

Like other existing mortality models, the time-varying factor model proposed in this paper can only extract linear features from the original data.  For future research, we are interested in developing a general non-parametric factor model, which can capture non-linear features in mortality data. Modern statistical techniques, such as neural networks and non-parametric estimation methods, can be used to fulfil this target. In addition, we are  interested in extending the age range considered in this research to include more advanced ages \citep{HUANG202095} and exploring time-varying factor models for multi-population mortality forecasting.

\medskip

\bibliography{timevarying}

\newpage
\section*{Appendix A}
Our analysis of the mortality forecasting in the main text focuses on age-specific  US mortality data of the total population. This appendix provides additional gender-specific results of the estimation and out-of-sample forecasting for both males and females.
\subsection*{Male}
\begin{figure} [hbt!]
	\centering
	\includegraphics[width=0.7\linewidth]{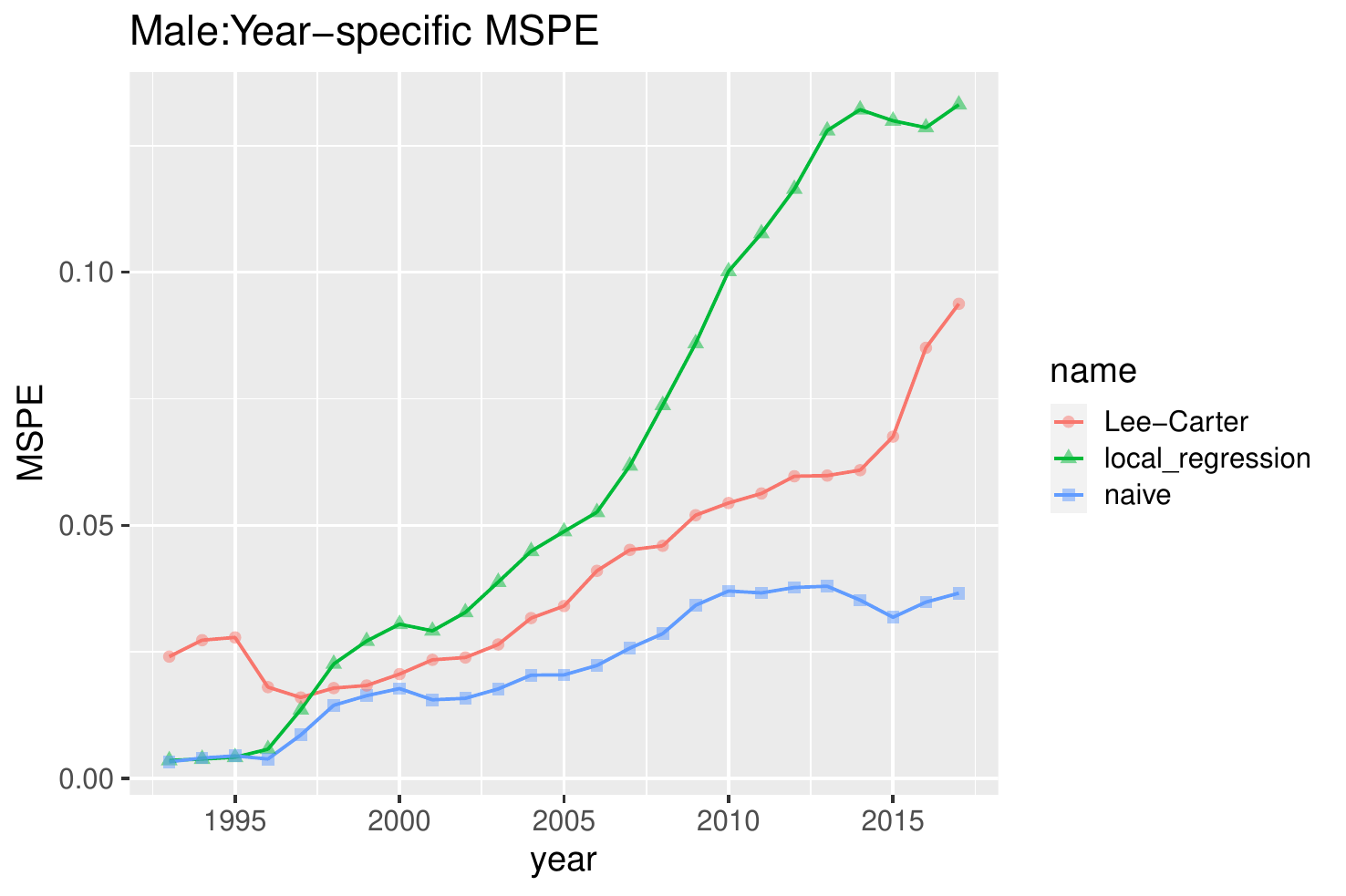}
	\caption{Year-specific MSPE for both the time-varying model and the Lee-Carter model; for time-varying model, both naive method and local regression method are used; \textbf{Male} subpopulation.
	}
	\label{a1}
\end{figure}

\begin{figure} [hbt!]
	\centering
	\includegraphics[width=0.7\linewidth]{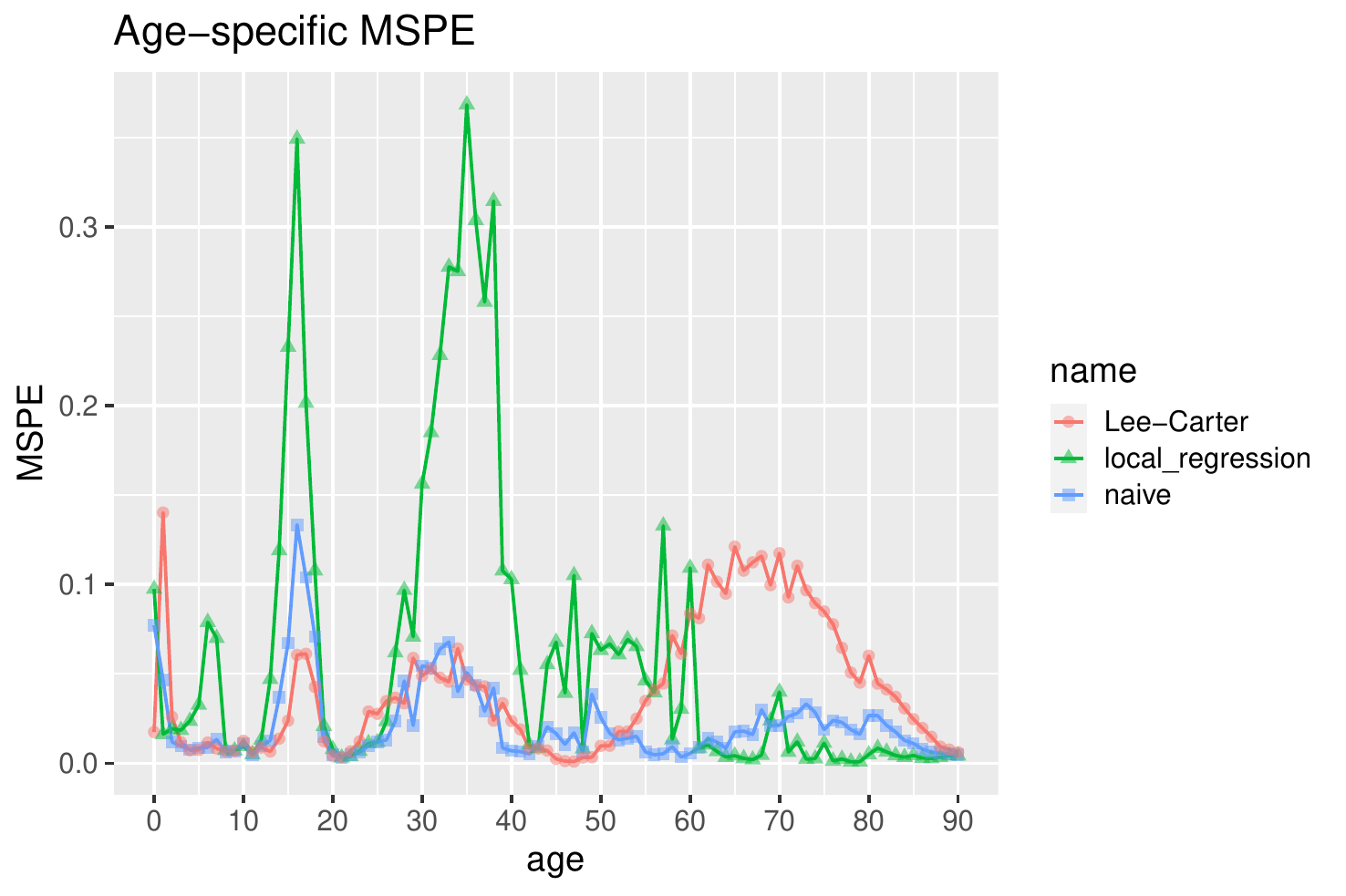}
	\caption{Age-specific MSPE for both the time-varying model and the Lee-Carter model; for time-varying model, both naive method and local regression method are used; \textbf{Male} subpopulation.
	}
	\label{a2}
\end{figure}

The results for the male subpopulation are showed in Figure \ref{a1} and Figure \ref{a2}.

In terms of the goodness of fit, the overall MSE for the Lee-Carter model is $0.008479$, while the MSE for the time-varying model is $0.002679$. From the perspective of the out-of-sample forecasting precision, the overall MSPE for the Lee-Carter model is $0.0412585$, while the MSPE for the time-varying model with the naive method is only $0.02247$, which is approximately $45\%$ less than the Lee-Carter model. Based on local linear regression, the MSPE of time-varying model is $0.06222$, which is larger than that for the Lee-Carter model. Thus, for the male subpopulation, compared with the Lee-Carter model, the prediction accuracy improves significantly by using the time-varying factor model based on the naive method. For the short-term forecasting, the naive method and local regression method have similar performance. However, for the long-term forecasting, the prediction accuracy of local regression method deteriorates as time goes by. For the age-specific forecasts, the time-varying model based on the naive method almost always obtains the best predictions. And roughly speaking, no matter which method we choose to use, the time-varying factor model is better at predicting the mortality rates of the older adulthood ($50\sim90$).

\subsection*{Female}
\begin{figure} 
	\centering
	\includegraphics[width=0.7\linewidth]{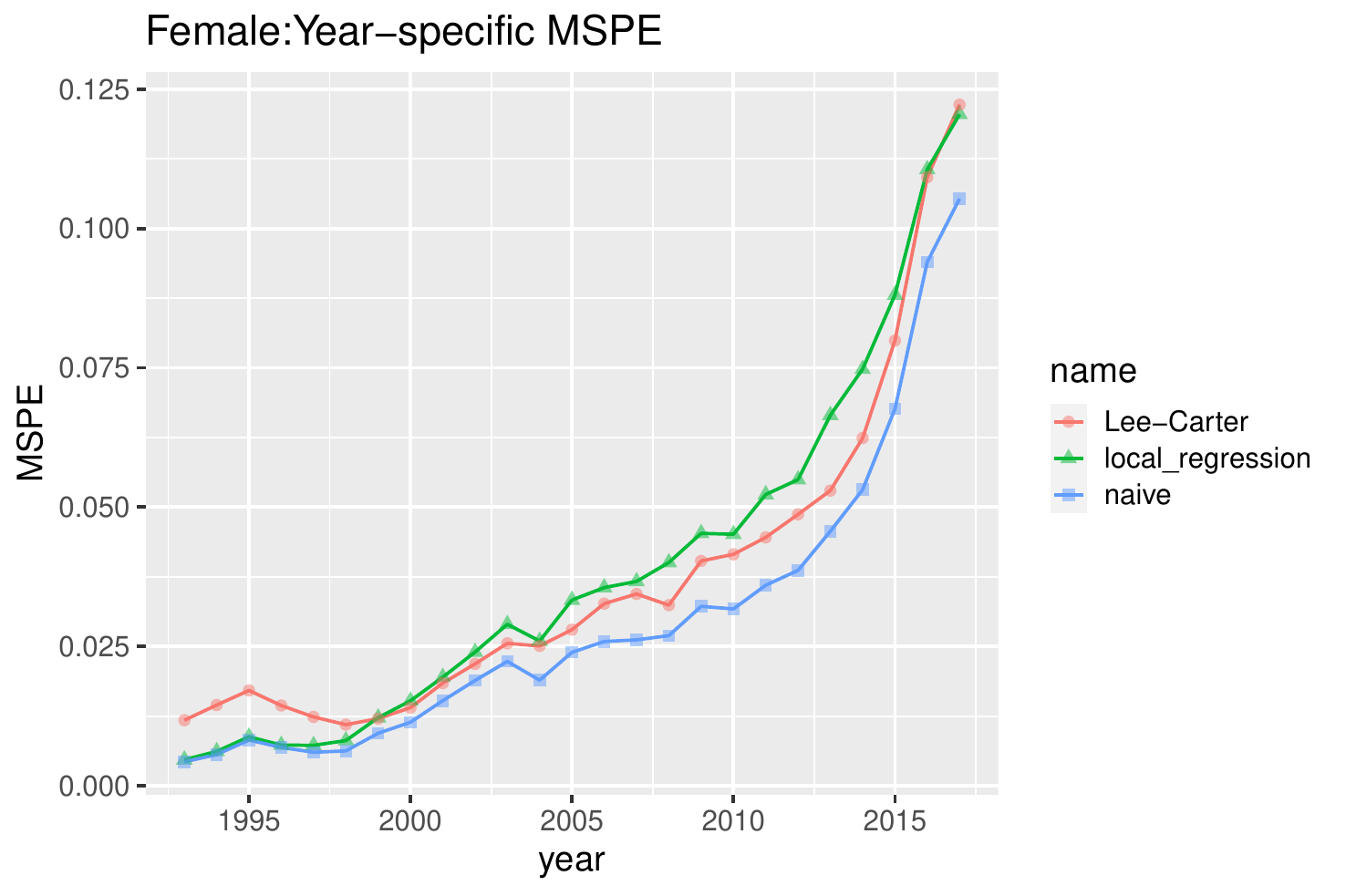}
	\caption{Year-specific MSPE for both the time-varying model and the Lee-Carter model; for time-varying model, both the naive method and local regression method are used; \textbf{Female} subpopulation.
	}
	\label{a3}
\end{figure}

\begin{figure} 
	\centering
	\includegraphics[width=0.7\linewidth]{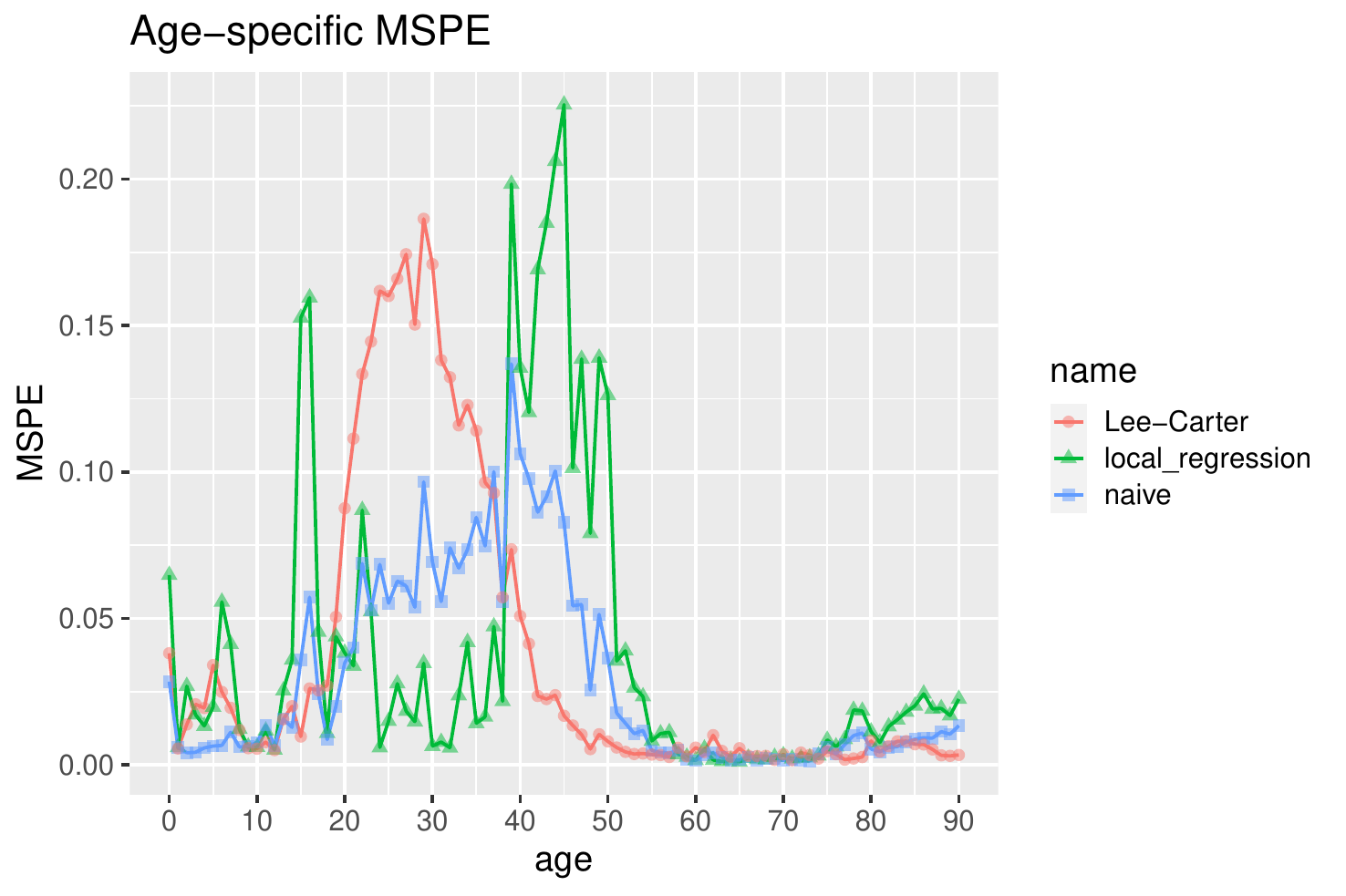}
	\caption{Age-specific MSPE for both the time-varying model and the Lee-Carter model; for time-varying model, both naive method and local regression method are used; \textbf{Female} subpopulation.
	}
	\label{a4}
\end{figure}

The results for the female subpopulation are showed in  Figure \ref{a3} and Figure \ref{a4}.

In terms of the goodness of fit, the overall MSE for the Lee-Carter model is $0.007337$, while the MSE for the time-varying model is $0.002061$. From the perspective of the out-of-sample forecasting precision, the overall MSPE for the Lee-Carter model is $0.03709$, while the MSPE for the time-varying model with the naive method is $0.02963$, which is $20\%$ less than the Lee-Carter model. Based on local linear regression, the MSPE of the time-varying model is $0.03887$, which is similar to (or slightly larger than) that for the Lee-Carter model. Thus, for the female subpopulation, compared with the Lee-Carter model, the prediction accuracy improves significantly by using the time-varying factor model based on the naive method. For the short-term forecasting, the naive method and local regression method have similar performance. However, for the long-term forecasting, the prediction accuracy of the local regression method deteriorates as time goes by. For the age-specific forecasts, the time-varying model based on the naive method almost always obtains the best predictions, except for the age group $40\sim55$. And roughly speaking, no matter which method we use, the time-varying factor model is better at predicting the mortality rates of the young children ($0\sim10$), young adulthood ($20\sim40$) and the older adulthood ($55\sim80$).

\section*{Appendix B}

\begin{figure}[H]
	\centering
	\begin{minipage}[c]{0.45\textwidth}
		\includegraphics[width=\linewidth, page=1]{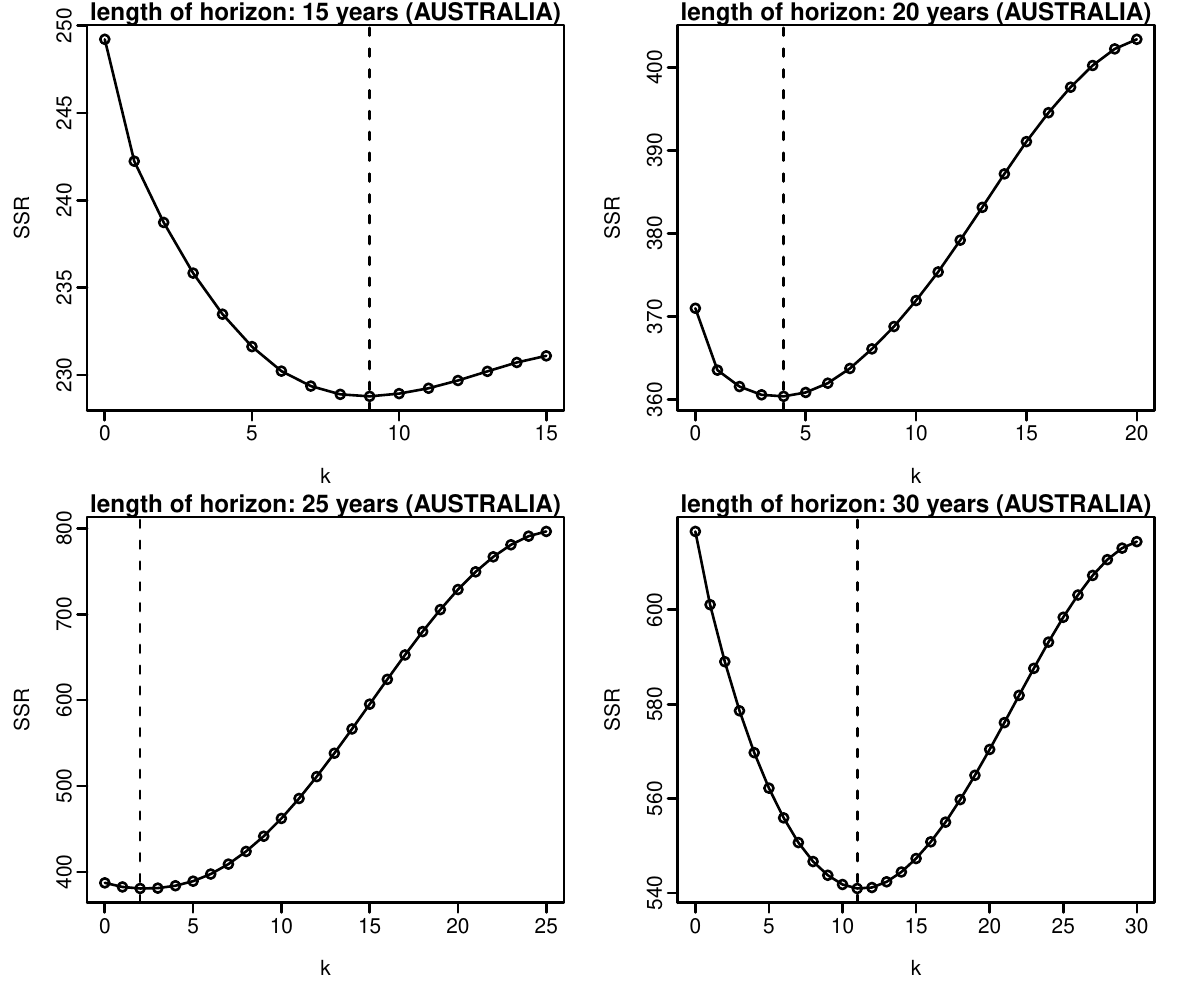}
		\includegraphics[width=\linewidth, page=3]{fig/a5.pdf}
		\includegraphics[width=\linewidth, page=5]{fig/a5.pdf}
	\end{minipage}
	\hspace*{0.05cm}
	\begin{minipage}[c]{0.45\textwidth}
		\includegraphics[width=\linewidth, page=2]{fig/a5.pdf}
		\includegraphics[width=\linewidth, page=4]{fig/a5.pdf}
		\includegraphics[width=\linewidth, page=6]{fig/a5.pdf}
	\end{minipage}
	\caption{Plots of the total sum of squared residuals (SSR) versus the length ($k$) of the short-term forecast horizon (based on the hybrid forecasting method of time-varying factor model); length of forecast horizon: $15$, $20$, $25$ and $30$}
	\label{a5}
\end{figure}

\section*{Appendix C}
In Section \ref{subsec: fit},  we only present the estimation results of the time-varying model with a single factor, as the number of factors for the US mortality data is estimated as 1. This appendix provides some further empirical results of fitting multi-factor time-varying model to the US mortality data. The number of factors is 6 (i.e. R = 6) in the following analysis. To solve the identification problem in the multi-factor context, we require the invertible matrix $\bbH_t$ as a diagonal matrix and normalize the $\bbb_{x, t}$ to sum to unity for each $t$. For more details, please refer to Section \ref{subsec: identification}.

\begin{figure} [hbt!]
	\centering
	\includegraphics[width=\linewidth]{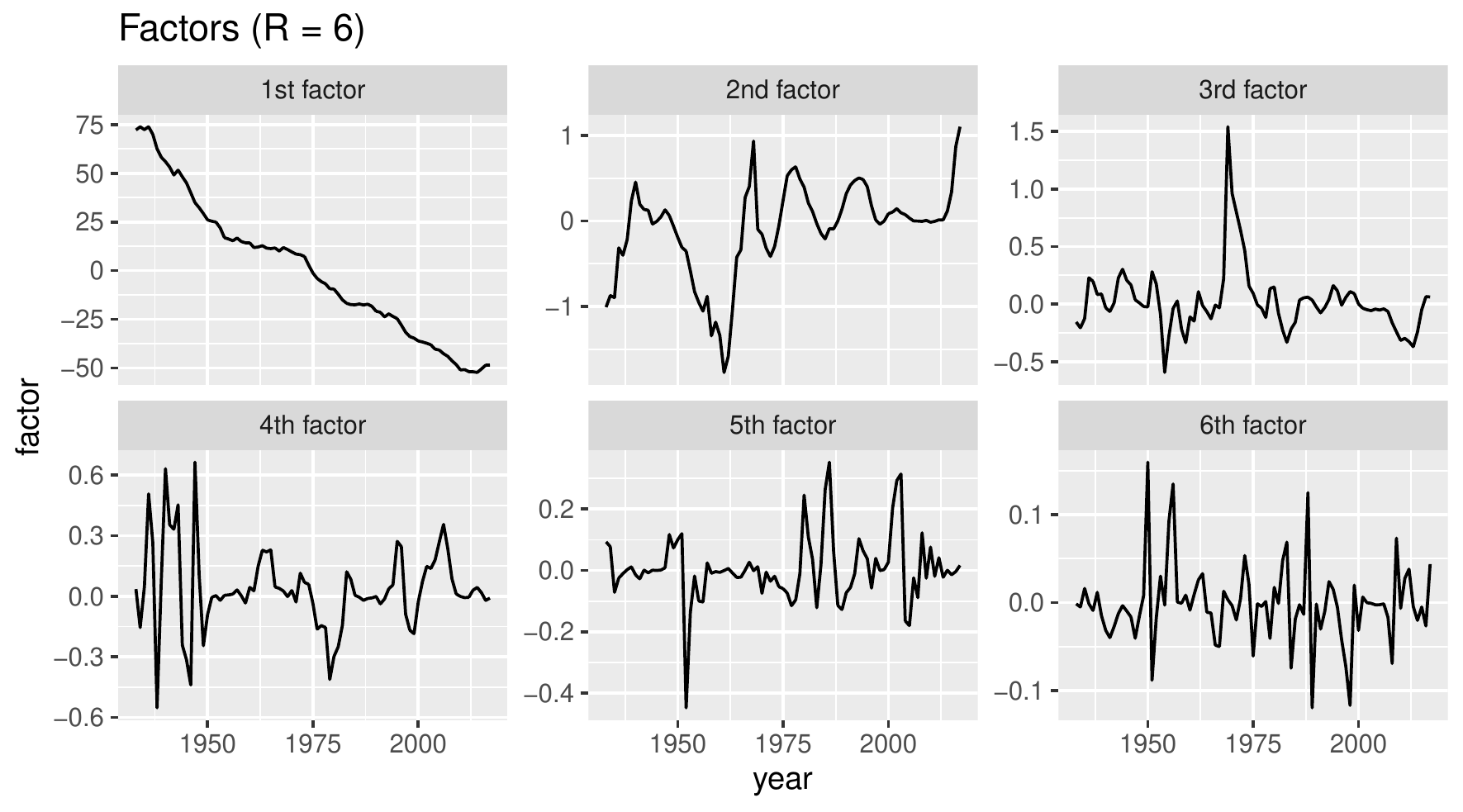}
	\caption{Plots of the estimated factors for the time-varying factor model ($R = 6$)
	}
	\label{r1}
\end{figure}

\begin{figure} [hbt!]
	\centering
	\includegraphics[width=0.8\linewidth, page=1]{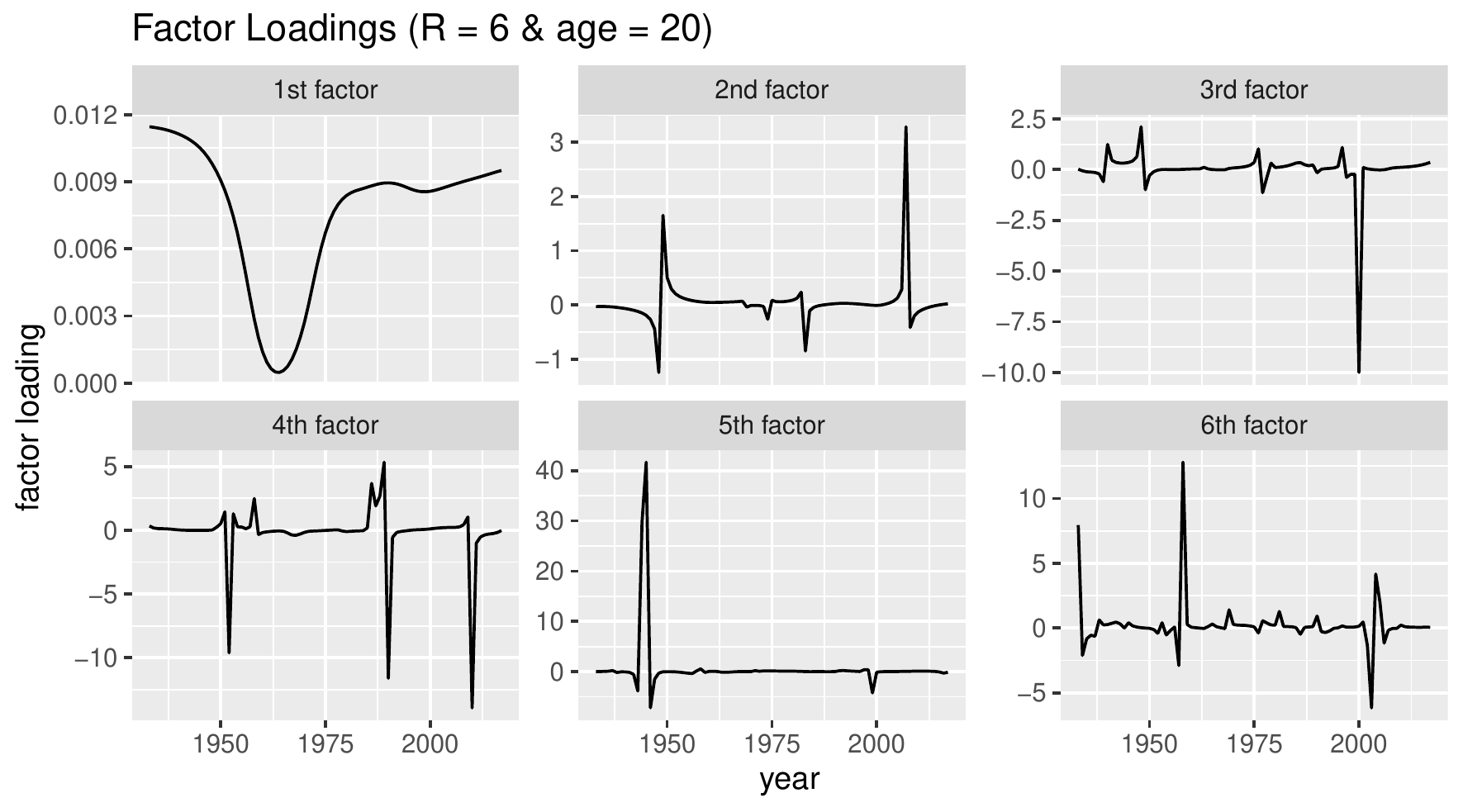}
	\includegraphics[width=0.8\linewidth, page=2]{fig/r2.pdf}
	\includegraphics[width=0.8\linewidth, page=3]{fig/r2.pdf}
	\caption{Plots of the estimated time-varying factor loadings for age 20, 40 and 60 ($R = 6$)
	}
	\label{r2}
\end{figure}

All six factors are depicted in Figure \ref{r1}. And in Figure \ref{r2}, we plot all the corresponding time-varying factor loadings for ages 20, 40 and 60. We can see that the first factor and its corresponding factor loadings remain the same as the results in Section \ref{subsec: fit}. The explanation is that theoretically, increasing the number of factors $R$ does not affect the estimation of the first common factor $\boldsymbol{\kappa}_t$ and the corresponding factor loading $\bbb_{x,t}$ in the localized PCA approach (see the estimation method in Section \ref{subsec: estimation}). The first factor captures the major trend of the mortality data, while the remaining factors are much less important with little information. It validates our decision to use one factor in the empirical analysis in Section \ref{sec: empirical}. 

\end{document}